\documentclass[%
 reprint,
 pra,
 floatfix,
 amsmath,amssymb,
 aps, widetext
]{revtex4-2}

\usepackage{graphicx} 
\usepackage{xspace}
\usepackage{enumerate}
\usepackage[dvipsnames]{xcolor}

\allowdisplaybreaks[2]

\begin{document}

\newcommand{\fe}{${}^{57}$Fe\xspace}



\newcommand{\ffc}{FFC\xspace}

\newcommand{\notheta}{\kappa}

\newcommand{\ler}{LER\xspace}

\newcommand{\expmatrix}{\mathcal{K}}

\title{A characterization and detection method for  x-ray excitation of M\"ossbauer nuclei beyond the  low-excitation regime}

\author{Lukas Wolff}
\affiliation{Max-Planck-Institut f\"ur Kernphysik, Saupfercheckweg 1, 69117 Heidelberg, Germany}

\author{J\"org Evers}
\affiliation{Max-Planck-Institut f\"ur Kernphysik, Saupfercheckweg 1, 69117 Heidelberg, Germany}

\date{\today}

\begin{abstract}
    Up to now, experiments involving M\"ossbauer nuclei have been restricted to the low-excitation regime. The reason for this is  the narrow spectral line width of the nuclei. This defining feature enables M\"ossbauer spectroscopy with remarkable resolution and convenient control and measurements in the time domain, but at the same time implies that only a tiny part of the  photons delivered by accelerator-based x-ray sources with orders-of-magnitude larger pulse bandwidth are resonant with the nuclei.
    X-ray free-electron lasers promise a substantial enhancement of the number of nuclear-resonant photons per pulse, such that excitations beyond the  low-excitation (\ler) regime come within reach. 
    This raises the question, how the onset of non-linear excitations could be experimentally verified. Here, we develop and explore a method to detect an excitation of nuclear ensembles beyond the \ler for ensembles of nuclei embedded in x-ray waveguides. It relies on the comparison of the x-rays coherently and  incoherently scattered off of the nuclei. As a key result, we show that the ratio of the two observables is constant within the \ler, essentially independent of the details of the nuclear system and the characteristics of the exciting x-rays. Conversely, deviations from this equivalence serve as a direct indication of excitations beyond the \ler. Building upon this observation, we develop a variety of experimental signatures both, for near-instantaneous impulsive and for temporally-extended non-impulsive x-ray excitation. Correlating coherently and incoherently scattered intensities further allows one to compare theoretical models of nonlinear excitations more rigorously to corresponding experiments.

\end{abstract}

\maketitle

\section{Introduction}

The extremely narrow line width of M{\"o}ssbauer transitions renders them ideal candidates for applications in precision spectroscopy and quantum optics~\cite{kalvius2012rudolf,yoshida2021modern,Roehlsberger2021}. Most commonly associated with the study of hyperfine interactions in solid state targets, recent years have witnessed a rising interest in M{\"o}ssbauer nuclei as an experimental platform for studying and controlling quantum dynamics and quantum optical effects~\cite{doi:10.1080/09500340.2012.752113,Roehlsberger2021,Rev_Roehlsberger_Evers,kuznetsova_quantum_2017,adams2019scientific,shv,Vagizov1990,Helistoe1991,PhysRevB.47.7840,PhysRevB.52.10268,Shvydko1996,PhysRevB.56.R8455,Hannon1999,Schindelmann2002,SmirnovPolariton2007,Roehlsberger2010,Roehlsberger2012EIT,Heeg2013SGC,Vagizov2014,Heeg2015SL,PhysRevLett.114.207401,haber_collective_2016,Heeg2017,Sakshath2017,Haber2017,bocklage2021.20097,Heeg2021,Shvydko2022,Heeg2022,PhysRevLett.129.213602}. This in part is facilitated by the experimental robustness and convenience of M{\"o}ssbauer setups that can often be operated at room temperature and ambient pressure in a solid-state target environment, and due to a time-separation of the nuclear signal from the much faster electronic background processes. Further progress is anticipated, e.g.,  due to advances in manufacturing specific nuclear environments~\cite{Roehlsberger2021} and due to the availability of new coherent high-brilliance x-ray sources \cite{EAA10,HXRSS12,SACLA,IOH19,EXFEL20,NMO21,SGG23} which  pave the way for new experiments based on coherence and tailored quantum optical and many-body properties of the nuclear system.

However, for many applications in quantum optics and spectroscopy, nonlinear light-matter interactions are essential.  
By contrast, experiments on M\"ossbauer nuclei have been restricted to the  low-excitation regime (\ler) so far, because of their narrow spectral line width which is orders of magnitude smaller than the bandwidth of x-ray pulses by state-of-the-art accelerator-based x-ray sources. 
This situation is expected to change with the recent availability of (seeded) x-ray-free electron lasers~\cite{EAA10,HXRSS12,SACLA,IOH19,EXFEL20,NMO21,SGG23}, which can provide a large number of nuclear-resonant photons per pulse. Further progress is anticipated with x-ray free-electron laser oscillators (XFELO)~\cite{PhysRevLett.100.244802,adams2019scientific}. Theoretical studies suggest that this may allow one to excite nuclear ensembles even up to the point of inversion \cite{Heeg_CollInv_2016,PhysRevResearch.4.L032007,RevModPhys.69.1085}. However, a first experiment on multi-photon excitation still found data consistent with linear excitation conditions \cite{chumakov_superradiance_2018}, and initially, future experiments are likely to only slightly surpass the \ler. While experiments to verify high excitation and inversion in nuclear resonant scattering have been suggested~\cite{Heeg_CollInv_2016}, experimentally relevant signatures at the onset of non-linear excitation of the nuclear ensemble are still largely lacking.

\begin{figure*}[t] 
\includegraphics[width=0.95\textwidth]{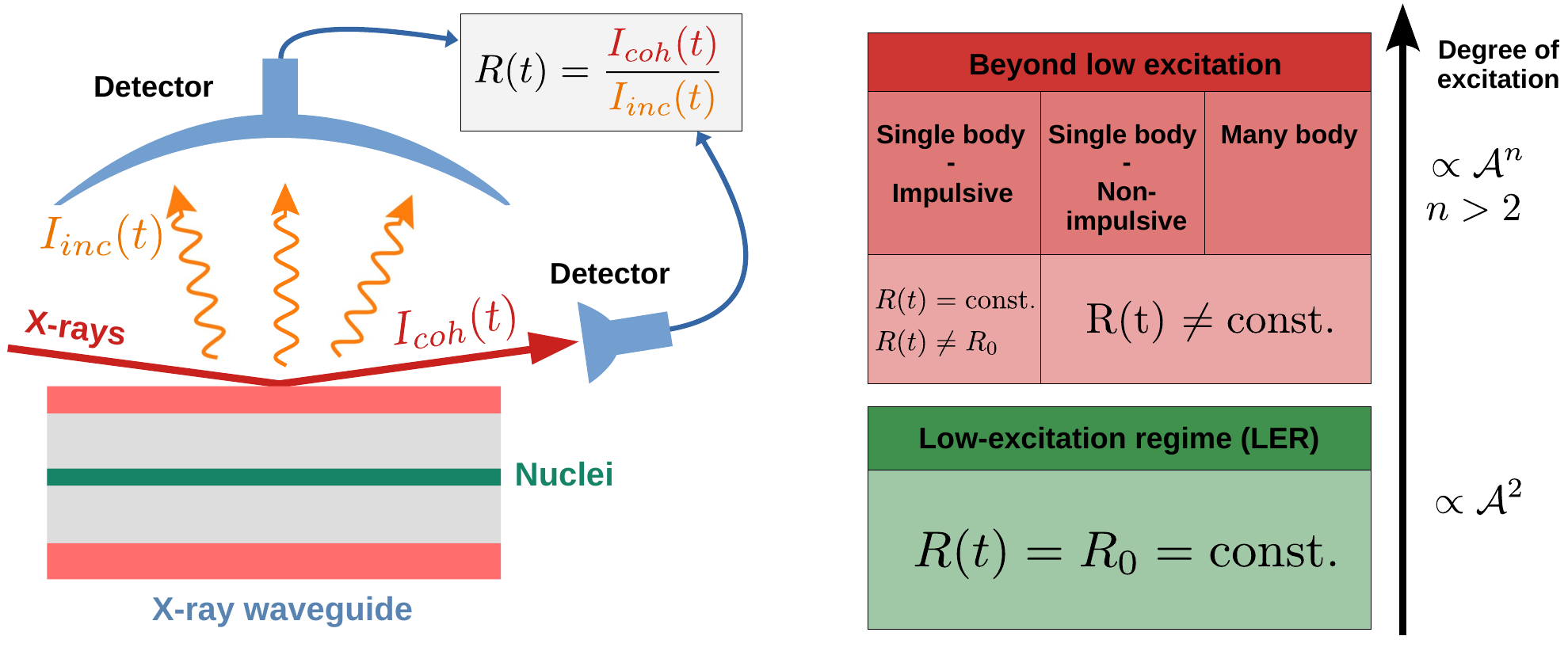}
\caption{Schematic setup and summary of the main results. The central goal of this work is to identify experimentally relevant signatures which enable one to verify the excitation of an ensemble of M\"ossbauer nuclei  beyond the low-excitation regime (\ler) explored up to now. For this, we consider an ensemble of two-level nuclei embedded in an x-ray waveguide, probed by near-resonant x-rays in grazing incidence. To detect the non-linear excitation, the coherently and the incoherently (e.g., following internal conversion) scattered intensities are recorded. Our theoretical analysis shows that in the \ler, the ratio $R(t)=R_0$ of the two intensities is constant, even though both intensities individually depend on time. We show this by analytically calculating the relevant dynamics of the excited-state populations and the x-ray-induced coherences of the general interacting $N$-body system in second order of the x-ray pulse area $\mathcal{A}$. Upon excitation beyond the \ler, the ratio $R$ changes its properties in a characteristic way. In case of near-instantaneous impulsive excitation of an effective single-particle system or a sufficiently weakly-coupled many-body system, the ratio $R$ remains constant, but changes its value from $R_0$ depending on the degree of excitation. 
For a strongly-interacting impulsively-driven many-body system, the ratio becomes time-dependent upon excitation beyond the \ler. In case of  non-impulsive x-ray excitation with duration of the order of the nuclear lifetime, the ratio is also time-dependent at higher excitation. Based on these results, a variety of different experimental signatures or data analysis approaches is developed which allow one to verify excitations beyond the \ler. \label{fig:main}}
\end{figure*}

To fill this gap, here, we develop and explore a method to verify the excitation of nuclear ensembles by intense x-ray light beyond the \ler. To this end, we propose to correlate two observables that are readily accessible in nuclear resonant scattering experiments, namely the highly directional coherently scattered intensity on the one hand and incoherent scattering products such as fluorescence emission and conversion electrons that are scattered into the entire solid angle on the other hand, see Fig.~\ref{fig:main}. Our results show that in the \ler, i.e. up to second order in the x-ray-nuclei interaction, these two observables become essentially equivalent. In particular, we  prove that the ratio of both observables becomes constant, a result that is largely independent of the details of the nuclear system and the temporal and spectral shape of the x-ray-nuclei interaction. Conversely, we demonstrate that already in leading  nonlinear excitation order, this ratio changes with the strength of the interaction and in a number of important cases becomes time-dependent. In particular, we study the case of impulsive excitation of nuclear ensembles with weak and strong nucleus-nucleus interactions corresponding to standard pulse conditions at accelerator-based light sources. Further, we identify clear experimental signatures of excitations beyond the \ler for near-monochromatic pulses, e.g., from synchrotron-M\"ossbauer-like sources~\cite{gerdau_nuclear_1985,PhysRevB.55.5811,PhysRevLett.102.217602,masuda_development_2008,potapkin_57fe_2012} generalized to operation at x-ray free-electron lasers~\cite{chumakov_superradiance_2018}. The correlation of coherently and incoherently scattered intensities also allows one to rigorously benchmark theoretical models of nonlinear excitations against experimental data and helps to  characterize deviations from effective low-excitation descriptions of nuclear ensembles.

The manuscript is structured as follows: In Sec.~\ref{sec:Theory} the theoretical model for the nuclear ensemble is introduced, the two relevant observables are presented and the cases of impulsive and non-impulsive x-ray excitation are defined. Sec.~\ref{sec:Signatures} outlines the basic principle of the distinction between the \ler and excitations beyond that regime. Subsequently the proof of the equivalence between coherent and incoherent dynamics up to second order in the x-ray-nuclei coupling is given first for effective two-level schemes and second for interacting many-body systems. The last two sections focus on signatures of excitation beyond the \ler for different pulse structures. Sec.~\ref{sec:impulsive} compares analytical results for weakly-coupled nuclear ensembles with numerical studies of strongly coupled nuclei upon impulsive excitations. Section~\ref{sec:non-impulsive} identifies different signatures for nonlinear excitation of effective two-level systems upon near-resonant and exponentially-decaying x-ray fields. Finally, Sec.~\ref{sec:summary} discusses and summarizes the results.

\section{Theoretical background \label{sec:Theory}}

\subsection{\label{sec:model}Theoretical model for the nuclear ensemble}
In the following, for definiteness, we focus our discussion on the case of nuclei embedded in planar thin-film waveguides, probed by the x-rays in grazing incidence on the waveguide structure~\cite{Hannon1999,Roehlsberger2021,RoehlsbergerBook}. Such photonic environments allow one to tailor the nuclear dynamics, and the possibility to enhance the nuclear excitation for a given x-ray pulse using a suitable design of the nuclear environment has been suggested~\cite{Heeg_CollInv_2016,PhysRevResearch.4.L032007,PhysRevA.105.013715,PhysRevA.106.053701}. Interestingly, the lossy nature of the x-ray waveguides  leads to an interplay of multiple cavity modes~\cite{PhysRevLett.130.263602} which may affect the nuclear dynamics favorably. Furthermore, one may expect that waveguides probed in reflection can be more stable under the action of intense x-ray light, as compared to thicker sample foils probed in forward direction. Finally, a detailed quantum optical description has been developed for the waveguide setting~\cite{Heeg2013Model,PhysRevA.91.063803,lentrodt_ab_2020,PhysRevA.102.033710}, which serves as the starting point for our present analysis.

In general, the interaction of x-rays with nuclei gives rise to a large variety of processes, e.g., based on recoil-less interaction or interaction with recoil, or on radiative or non-radiative de-excitation channels. A detailed discussion of these contributions in the LER regime can be found, e.g., in~\cite{SmirnovPolariton2007,Hannon1999}. In the following, we aim at a description of the nuclear dynamics beyond the LER, focusing on coherent scattering in propagation direction of the driving x-ray pulse and incoherent emission following internal conversion as the main observables. To this end, we start by modelling the nuclei as a generic ensemble of $N$ identical interacting two-level systems using the Hamiltonian~\cite{Agarwal1974,Ficek_Swain,Kiffner_Vacuum_Processes,lentrodt_ab_2020,PhysRevA.102.033710,Longo2016,PhysRevLett.112.193601,PhysRevA.90.063834,Asenjo-Garcia2017}
\begin{align} \label{eq:ManyBodyHam}
\hat{H} =& \hbar \sum_{n=1}^N  \omega_0 \: \hat{\sigma}^+_n \hat{\sigma}^-_n - \frac{\hbar}{2} \sum_{n=1}^N \left[\Omega(\mathbf{r}_n,t) \,\hat{\sigma}^+_n +h.c.\right] \nonumber \\[1ex]
&-\hbar \sum_{n,n'=1}^N  J_{nn'} \: \hat{\sigma}^+_n \hat{\sigma}^-_{n'}\,.
\end{align}
Here, $n$ and $n'$ label the individual two-level systems and $\hat{\sigma}_n^{\pm}$ are the raising and lowering operators of nucleus $n$ in its two-level Hilbert space. Further, $\omega_0$ denotes the nuclear transition frequency and
\begin{align}
\Omega(\mathbf{r}_n,t) = \frac{\mathbf{d}\, \mathbf{E}(\mathbf{r}_n,t)}{\hbar}
\end{align}
the semi-classical light-matter coupling in the form of the time-dependent Rabi frequency with $\mathbf{E}(\mathbf{r}_n,t)$ describing the electric field amplitude at the position $\mathbf{r}_n$ of nucleus $n$, and $\mathbf{d}$ the transition dipole moment~\footnote{Note that the M{\"o}ssbauer transitions in some isotopes such as ${}^{57}$Fe may instead feature other dominant multipole moments, such as M1 magnetic dipole moments. In this case, the expression for the Rabi frequency should be modified accordingly}.
Finally, possible interactions between the nuclei are included via the dipole-dipole coupling parameters $J_{nn'}$ which satisfy the symmetry property $J_{nn'} = J^*_{n'n}$.

The nuclear many-body dynamics is then characterized by a density operator $\hat{\rho}^{\textrm{NB}}$ governed by the master equation~\cite{Agarwal1974,Ficek_Swain,Kiffner_Vacuum_Processes,lentrodt_ab_2020,PhysRevA.102.033710,Longo2016,PhysRevLett.112.193601,PhysRevA.90.063834,Asenjo-Garcia2017}
\begin{align} \label{eq:Master}
\frac{d}{dt}\hat{\rho}^{\textrm{NB}} = \frac{1}{i\hbar}\left[\hat{H}, \hat{\rho}^{\textrm{NB}}\right] + \mathcal{L}[\hat{\rho}^{\textrm{NB}}] \, ,
\end{align}
where ``$\textrm{NB}$'' stands for N-body, and the Lindblad term is given by
\begin{align} \label{eq:ManyBodyLind}
\mathcal{L}[\hat{\rho}^{\textrm{NB}}] =& \sum_{n ,n'=1}^N\Gamma_{nn'}\left(2\hat{\sigma}^{-}_{n'}\hat{\rho}^{\textrm{NB}}\hat{\sigma}^{+}_{n}-\left\{\hat{\sigma}^{+}_n\hat{\sigma}^-_{n'},\hat{\rho}^{\textrm{NB}}\right\}\right) \nonumber \\
&+ \sum_{n=1}^N\Gamma_{\mathrm{IC}}\left(2\hat{\sigma}^{-}_n\hat{\rho}^{\textrm{NB}}\hat{\sigma}^{+}_{n}-\left\{\hat{\sigma}^{+}_n\hat{\sigma}^-_{n},\hat{\rho}^{\textrm{NB}}\right\}\right)  \,.
\end{align}
It incorporates both, single-particle decay as diagonal elements with $n=n'$, and incoherent dipole-dipole couplings between the nuclei with $n \neq n'$. Note that the total natural line width $\gamma$ of the nuclei comprises radiative decay contributions ($\propto \Gamma_{nn}$), and non-radiative internal conversion contributions ($\propto \Gamma_{\mathrm{IC}}$), 
\begin{align}
    \gamma = 2\left(\Gamma_{nn} + \Gamma_{\mathrm{IC}} \right)\,.
\end{align}

The coupling constants entering the master equation Eq.~(\ref{eq:Master}) can conveniently be calculated ab-initio using the classical Green's function~\cite{Gruner1996,Dung2002,Scheel2008,Asenjo-Garcia2017,NovotnyHecht2006,Buhmann2007} characterizing the nuclear environment~\cite{lentrodt_ab_2020}, which is analytically known~\cite{Tomas1995,Buhmann2007,Asenjo-Garcia2017}.  In turn, a suitable optimization of the environment can be used to reverse engineer desired couplings~\cite{Bennett_2020,PhysRevA.105.013715,PhysRevA.106.053701}. 

The many-body problem Eq.~(\ref{eq:Master}) in general is challenging to solve. However, in the \ler, enforced in the equations of motion by neglecting possible populations of the nuclear excited states by setting $\langle \hat{\sigma}^+_n \hat{\sigma}^-_n\rangle=0$, the problem allows for a substantial reduction of the relevant Hilbert space. In this case, by rewriting the system in Fourier space in terms of a spin-wave basis, the problem of many interacting nuclei embedded in the cavity environment can equivalently be rewritten in terms of an effective single-particle level scheme~\cite{Roehlsberger2010,Roehlsberger2012EIT,Heeg2013Model,PhysRevA.91.063803,lentrodt_ab_2020,Rev_Roehlsberger_Evers,Roehlsberger2021,PhysRevA.102.033710,PhysRevA.106.053701,PhysRevA.105.013715}.
Interestingly, the effective level scheme may differ from the original level scheme of the individual nuclei. The number of relevant energy eigenstates can be engineered, and it may also comprise additional couplings between levels induced by the cavity environment which can simulate otherwise unavailable control laser fields. As a result, level schemes can be realized which otherwise are not available with M{\"o}ssbauer nuclei.  This feature forms the basis of most experiments on nuclear quantum optics with nuclei in waveguides reported so far~\cite{Rev_Roehlsberger_Evers,Roehlsberger2021,Roehlsberger2010,Roehlsberger2012EIT,Heeg2013SGC,Heeg2015SL,haber_collective_2016,Haber2017,PhysRevLett.114.207401,PhysRevLett.129.213602}

It is expected that this equivalent description in terms of a single few-level system breaks down towards higher excitation of the nuclear ensemble~\cite{lentrodt_ab_2020,PhysRevLett.112.193601,PhysRevA.90.063834}. Nevertheless, the single-particle description provides a good starting point for the following analysis of experimental signatures at the onset of effects beyond the \ler. Afterwards, in Sec.~\ref{sec:PerturbInt}, we will also consider the full many-body dynamics in leading and next-to-leading order of the interaction between x-rays and nuclei accompanied by numerical simulations of higher excitation orders to explore possible deviations from the single-particle results in the excitation beyond the linear regime.

Note that a similar treatment in principle can also be applied to nuclear forward scattering (for an introduction, see e.g. \cite{Hannon1999,RoehlsbergerBook}), by employing the corresponding free-space Green's function to calculate the parameters entering the master equation. However, in this geometry, the incident x-rays typically excite multiple eigenmodes of the many-body Hamiltonian, and  propagational effects arising due to multiple interactions in the thicker samples lead to further modifications of the scattered light signatures~\cite{PhysRev.120.513,Hannon1999}. Therefore, for simplicity, we focus on reflection geometries in the following analysis.

\subsection{\label{sec:observables}Observables}

In the following discussion, we will consider two standard observables in nuclear resonant scattering, see Fig.~\ref{fig:main}. First, the time-dependent intensity $I_{coh}(t)$ of the coherently scattered  x-rays. This quantity is highly directional: In forward scattering geometry, this signature is emitted in forward direction, due to interference between the scattering contributions of the different nuclei~\cite{Smirnov1986,Hannon1999,RoehlsbergerBook}. In reflection geometry, it is emitted in a direction essentially given by Bragg's law.  Second, the time-dependent intensity $I_{inc}(t)$ of the incoherent signatures, e.g., fluorescence photons or conversion electrons of the nonradiative de-excitation of the nuclei via internal conversion~\cite{PhysRevB.53.171,PhysRevB.49.1513,Baron_1996,PhysRevB.73.024203,SmirnovPolariton2007}. The relative contribution of the nonradiative to the radiative channel to the total nuclear decay is described  by the internal conversion coefficient $\alpha$ \cite{Hannon1999,RoehlsbergerBook}. Note that the two observables can be measured concurrently~\cite{SmirnovPolariton2007}. In the following, we will show that the comparison of these two observables allows one to identify excitations of the nuclei beyond the LER.

In incoherent scattering, the nuclei decay independently, such that the observed signal intensity is  proportional to the sum of excited-state populations of the nuclei~\cite{Agarwal1974,Ficek_Swain},
\begin{align}
I_{inc}(t) \propto \sum_{n=1}^N \: \langle \hat \sigma^+_n \hat \sigma^-_n\rangle \,. \label{eq:Iinc} 
\end{align}
Note that the pre-factors will not be of relevance in the following analysis. This  has the additional advantage that experimental details such as the detection geometry, or attenuation within the sample or between sample and detector do not have to be characterized quantitatively for our analysis. 

In contrast, the radiatively emitted scattered light can be evaluated by relating the positive- and negative-frequency components of the electric field operators $\hat E^{(\pm)}$ to the transition operators $\hat \sigma^\mp_n$~\cite{Agarwal1974,Ficek_Swain}, which act as source operators for the emitted radiation. The total coherently emitted intensity in direction $\mathbf{k}_{out}$ can then be written as~\cite{Agarwal1974,Ficek_Swain},
\begin{align}
I_{rad}(t, \mathbf{k}_{out}) \propto \sum_{n,m=1}^N \langle \hat{\sigma}^+_n\hat{\sigma}^-_m\rangle\: e^{i\mathbf{k}_{out}(\mathbf{r}_n-\mathbf{r}_m)}\,,
\end{align}
where $\mathbf{k}_{out}$ is the wave vector of the emitted radiation, and we again have omitted the pre-factors. Using a decomposition of the transition operators into their expectation values and a fluctuation part $\hat\sigma^\pm_n = \langle \hat\sigma^\pm_n \rangle + \delta \hat\sigma^\pm_n$~\cite{cct}, we can extract the coherently scattered contribution as
\begin{align}
I_{coh}(t, \mathbf{k}_{out}) &\propto \sum_{n,m=1}^N \langle \hat{\sigma}^+_n\rangle \,\langle \hat{\sigma}^-_m\rangle\: e^{i\mathbf{k}_{out}(\mathbf{r}_n-\mathbf{r}_m)}\nonumber \\
&= \left\lvert \sum_{n=1}^N e^{i\mathbf{k}_{out}\mathbf{r}_n}\langle \hat{\sigma}^+_{n}\rangle \right\rvert^2 \,. \label{eq:Icoh}
\end{align}
The last expression clearly exhibits the interference between the contributions scattered by the individual two-level nuclei. If the incident x-rays with wave vector $\mathbf{k}_{in}$ imprint a position-dependent phase pattern on the two-level nuclei, the additional phase accumulated due to $\mathbf{k}_{out}$ together with the sum over all two-level atoms leads to the directional emission described at the beginning of this section.

Note that the incoherently and coherently scattered intensities Eqs.~(\ref{eq:Iinc}),~(\ref{eq:Icoh}) can be expressed in terms of single-particle reduced density matrix elements,
\begin{align}
\langle \hat{\sigma}^+_n\hat{\sigma}^-_n\rangle =&\textrm{Tr}\left[\hat{\rho}^{\textrm{NB}}\hat{\sigma}^+_n\hat{\sigma}^-_n\right]  = \rho_{e_ne_n}\,, \\[1ex]
\langle \hat{\sigma}^+_n\rangle = &\textrm{Tr}\left[\hat{\rho}^{\textrm{NB}}\hat{\sigma}^+_n\right]=\rho_{g_ne_n}\,.
\end{align}
Here, $\lvert g_n\rangle$ [$\lvert e_n\rangle$] denote the ground [excited] state of nucleus $n$, $\hat{\rho}^{\textrm{NB}}$ is the N-body density matrix the dynamics of which is governed by the master equation Eq.~(\ref{eq:Master}), and $\textrm{Tr}\left[\cdot\right]$ denotes the trace over the many-body Hilbert space. Analogous relations hold for effective single-particle level schemes as introduced below in Sec.~\ref{sec:model}.

\subsection{\label{sec:x-ray}Impulsive and non-impulsive x-ray excitation}

Throughout this work, we consider two qualitatively different x-ray excitation approaches for the nuclei. Accelerator-based x-ray sources typically deliver x-ray pulses with durations on the $\sim$ps (synchrotron) or $\sim$fs (x-ray free electron laser) scale. In contrast, typical lifetimes of standard M\"ossbauer isotopes are orders of magnitude longer  (for example, the natural lifetime of the most commonly used M\"ossbauer resonance in ${}^{57}$Fe is $141$~ns~\cite{RoehlsbergerBook}). As a result, the x-ray excitation is {\it impulsive} in the sense that it is near-instantaneous as compared to all natural time scales of decay and coupling dynamics  of the nuclei. Therefore, collective effects such as couplings between the nuclei or their decay processes can be completely neglected throughout the x-ray excitation, and the nuclear excitation dynamics can be evaluated simply by considering the x-ray induced dynamics on the single-nucleus level. After initial excitation, the nuclei then evolve on their natural time scales in the absence of the exciting x-ray pulse. This temporal separation of excitation and subsequent nuclear ensemble dynamics  considerably simplifies the analysis.

Next to this impulsive excitation, we further consider the case in which the duration of the incident x-ray field is not restricted to very short times.  We denote this more general case as {\it non-impulsive} excitation, and will in particular consider the case in which the duration of the driving x-ray field is comparable to the other evolution time scales of the nuclei. This situation becomes of relevance if x-ray pulses are used which are monochromatic on the nuclear energy scales, e.g., delivered by a synchrotron (or analogously extended free-electron laser) M\"ossbauer source~\cite{gerdau_nuclear_1985,PhysRevB.55.5811,PhysRevLett.102.217602,masuda_development_2008,potapkin_57fe_2012,chumakov_superradiance_2018}, or in setups employing additional reference absorbers to shape the incoming x-ray pulse like the recently demonstrated coherent control schemes for nuclear dynamics~\cite{Heeg2021}. In the non-impulsive case, the complete dynamics involving x-ray excitation, couplings and decay processes must be considered at the same time.

In Sections~\ref{sec:impulsive} and \ref{sec:non-impulsive}, the two cases will be analyzed separately.

\section{Characterization and detection of dynamics beyond the low-excitation regime \label{sec:Signatures}}

\subsection{\label{sec:simple-example}Example illustrating the approach to detect excitation beyond the low-excitation regime}

As discussed in Sec.~\ref{sec:model}, effective single-particle models for nuclei in waveguides provide a good starting point for the analysis of experimental signatures at the onset of effects beyond the \ler. Here, we start with the simplest possible case, and consider the excitation dynamics of a single two-level system  resonantly driven by an impulsive x-ray pulse. 

We denote the ground and excited states of the two-level system as $|g\rangle$ and $|e\rangle$, respectively, and characterize its state via a density operator with matrix elements $\rho_{ij}(t)$ ($i, j\in \{e,g\}$; see Appendix~\ref{app:SelfDerive} for the equations of motion).
Assuming the system to be initially in its ground state, i.e. 
\begin{align}
    \rho_{ee}(t=0) &= 0\,, \qquad  \rho_{ge}(t=0) = 0\,, \nonumber\\
    \rho_{eg}(t=0) &= 0\,, \qquad \rho_{gg}(t=0) = 1\,,  \nonumber
\end{align}
the time-dependent  density matrix elements describing the excited-state population and the x-ray induced coherence in a suitable interaction picture can be evaluated using the area theorem~\cite{AllenEberly1975, MeystreSargent2007} to give
\begin{subequations}
\label{eq:PulseAreaNoInit}
\begin{align} 
\rho_{ee}(t) &= \sin^2\left[\mathcal{A}(t)\right] \,,  \\
\rho_{ge}(t) &= -\frac{i}{2}e^{i\phi}\sin\left[2\mathcal{A}(t)\right]\,,
\end{align}
\end{subequations}
where $t$ denotes the time after the impulsive x-ray pulse has passed the system, which is much shorter than the lifetime of the resonance such that the decay can be neglected. The x-ray pulse area $\mathcal{A}$ is given by 
\begin{align} \label{eq:Area}
\mathcal{A}(t) = \frac{1}{2}\int^{t}_{t_0} \,|\Omega(t')|\, dt'\,,
\end{align}
where $\Omega(t) =|\Omega(t)|\,\exp(i\phi) $ is the Rabi frequency in the interaction picture proportional to the time-dependent x-ray field amplitude, with constant phase $\phi$ which accounts, e.g., for the spatial dependence of the incident x-ray field. 
Note that the pulse area enters the off-diagonal density matrix element corresponding to the x-ray induced  coherence with a pre-factor of $2$. As a result, the two quantities $\rho_{ee}(t)$ and $\lvert \rho_{ge}(t) \rvert^2$ related to the incoherent and coherent x-ray emission from the nuclei are not equivalent in general. 

To explore the relation of the two observables in more detail, we expand Eqs.~(\ref{eq:PulseAreaNoInit}) for the case of low pulse area,
\begin{subequations}
\begin{align}
\rho_{ee}(t) &= \mathcal{A}(t)^2 - \frac{\mathcal{A}(t)^4}{3} + \frac{2\mathcal{A}(t)^6}{45}  +\dots\,, \\
|\rho_{ge}(t)|^2 &= \mathcal{A}(t)^2 - \frac{4\mathcal{A}(t)^4}{3} + \frac{32\mathcal{A}(t)^6}{45}+\dots \,. 
\end{align}
\end{subequations}
We find that, in leading order, the coherently and incoherently scattered light proportional to the coherence squared and the population are equivalent,
\begin{align} \label{eq:AreaId}
\rho_{ee}^{(0-2)}(t) = \mathcal{A}^2(t) = \lvert \rho_{ge}^{(0-2)}(t) \rvert^2\,,
\end{align}
where the superscript $(0-2)$ indicates the Taylor expansion including all contribution up to second order of the indexed quantities.

In contrast, if the nuclei are excited beyond the leading low-excitation order, we find that the two observables differ, 
\begin{align} \label{eq:4thOrder}
\rho^{(4)}_{ee}(t) \neq \left(|\rho_{ge}(t)|^2\right)^{(4)} = 2\textrm{Re}\left[\rho^{*(1)}_{ge}(t)\rho^{(3)}_{ge}(t)\right]\,,
\end{align}
where the superscript $(i)$ denotes the $i$th-order contribution of the series expansion only. 
As a result, we conclude that suitably analyzed deviations in the two observables provide a direct signature for the excitation of the nuclear ensemble beyond the \ler.

Equation~(\ref{eq:4thOrder}) also shows how the expansion of the off-diagonal density matrix element itself enters the expansion of its magnitude squared. As expected, the leading order violating the equivalence of the coherent and incoherent scattering comprises contributions from the third order of the off-diagonal density matrix elements, which illustrates the significance of the x-ray-nuclei interaction beyond the linear regime for this contribution.

In analyzing experimental data, it may be favorable to consider ratios of coherently and incoherently scattered light, corresponding to suitable ratios of squared coherences and populations such as $|\rho_{ge}|^2/\rho_{ee}$ as experimentally-accessible quantities to characterize excitation beyond the leading low-excitation order. The reason is that then, experimental aspects such as pre-factors related to the detection geometry or efficiency become largely irrelevant in analyzing the data. The ratio expands in orders of the pulse area as
\begin{align}
\frac{|\rho_{ge}(t)|^2}{\rho_{ee}(t)} &= \frac{\mathcal{A}(t)^2 - \frac{4\mathcal{A}(t)^4}{3} + \frac{32\mathcal{A}(t)^6}{45}+\dots}{\mathcal{A}(t)^2 - \frac{\mathcal{A}(t)^4}{3} + \frac{2\mathcal{A}(t)^6}{45}  +\dots}\nonumber \\
&= 1 - \mathcal{A}^2(t) + \frac{\mathcal{A}^4(t)}{3} + \dots\,.
\end{align}
Thus, it deviates from unity already in second order of the pulse area. However, it is important to note that this is due to cancellations in the expansion order of the numerator and the denominator of the ratio. Corrections of order $\mathcal{A}^2(t)$ in the ratio may only occur if the off-diagonal density matrix elements $\rho_{eg}$ and $\rho_{ge}$ have contributions of order $\mathcal{A}^3(t)$ or higher, and/or the populations $\rho_{ee}$ of order $\mathcal{A}^4(t)$ or higher. Therefore, we can attribute deviations in the ratio from unity to excitations of the system beyond the \ler.

In the following Sections, we will extend the characterization of excitation beyond the \ler based on the relation between coherently and incoherently scattered radiation to more complex settings and to non-impulsive nucleus-field interactions, and will develop various approaches to analyze the two observables for this purpose.

\subsection{Effective two-level system excited by non-impulsive x-ray fields}

In Sec.~\ref{sec:simple-example} we have used the simplest case of the excitation dynamics of a single effective two-level system driven by an impulsive x-ray field to illustrate the main idea of comparing the coherent and the incoherent light scattering off of the nuclei to identify excitation beyond the \ler. Next, we develop this argument further by deriving a self-consistent solution to the dynamics of the two-level system driven by  x-ray fields with arbitrary time-dependence and including decay dynamics. This will allow us to also study the non-impulsive x-ray excitation case. Note that the effective level scheme may have decay rates or transition frequencies which differ from the bare nuclear properties~\cite{Roehlsberger2010,Heeg2013Model,lentrodt_ab_2020,PhysRevA.105.013715}. Nevertheless, for notational simplicity, in the following, we will continue to use  the symbols $\gamma$ and $\omega_0$ introduced above as the single-nucleus properties also in the effective two-level case.

\subsubsection{Self-consistent solution for the effective two-level system}

As discussed in Sec.~\ref{sec:model}, the nuclear many-body system in a waveguide can be modeled using an effective single-particle description in the \ler. For the simplest case of an effective two-level scheme, self-consistent solutions to the equations of motion for the excited-state population and the x-ray-induced coherence can be derived, which are given by (see Appendix~\ref{app:SelfDerive} for the derivation)
\begin{subequations}
\label{eq:SelfConsNoInit}
\begin{align}  
    &\rho_{ee}(t,t_0)= \frac{1}{2}e^{-\gamma t}\mathrm{Re}\Big[\int^{t}_{t_0}dt'e^{\frac{\gamma}{2} t'}e^{-i\omega_0 t'}\Omega^{*}(t') \times \nonumber \\
    &\qquad \qquad  \times \int^{t'}_{t_0}dt''e^{i\omega_0 t''}e^{\frac{\gamma}{2} t''}\Omega(t'')\Big] \nonumber \\
    &\qquad  -e^{-\gamma t}\mathrm{Re}\Big[\int^{t}_{t_0}dt'e^{\frac{\gamma}{2} t'}e^{-i\omega_0 t'}\Omega^{*}(t') \nonumber \\
    & \qquad \qquad \times\int^{t'}_{t_0}dt''e^{i\omega_0 t''}e^{\frac{\gamma}{2} t''}\Omega(t'')\rho_{ee}(t'',t_0)\Big] \,, \label{eq:SelfPopNoInit} \\
     &\rho_{ge}(t,t_0)= -\frac{i}{2}e^{i\omega_0 t}e^{-\frac{\gamma}{2} t}\int^{t}_{t_0}dt'e^{-i\omega_0 t'}\Omega^*(t')e^{\frac{\gamma}{2} t'} \nonumber \\
    &\qquad  -ie^{i\omega_0 t}e^{-\frac{\gamma}{2} t}\int^{t}_{t_0}dt'e^{-i\omega_0 t'}e^{-\frac{\gamma}{2} t'}\Omega^*(t') \times \nonumber \\
    &\qquad \qquad \times \int^{t'}_{t_0}dt''e^{\gamma t''}\mathrm{Im}\big[\Omega (t'')\rho_{ge}(t'',t_0)\big] \,. \label{eq:SelfCohNoInit}
\end{align}
\end{subequations}
Here, we have assumed that the effective nucleus is initially in the ground state at time $t_0$. Note that as expected, the population only comprises terms of even orders in the nucleus-field coupling $\Omega(t)$ while the coherence only comprises odd orders.

In the next Subsection~\ref{sec:ProofCorrespond}, we will systematically expand this solution in orders of the driving x-ray field amplitude, in order to establish the relations between the excited-state population and the x-ray induced coherences which will allow us to identify signatures for excitation beyond the \ler.

\subsubsection{Coherence-population correspondence in second-order of the driving x-ray field \label{sec:ProofCorrespond}}

As illustrated in Sec.~\ref{sec:simple-example}, our approach to identify x-ray excitations of the nuclear ensemble beyond the \ler relies on a comparison of the coherently and incoherently scattered light, which relate to the nuclear excited-state population and the x-ray-induced coherence squared. It is based on the result that the population and the absolute value of the coherence squared are identical up to second order in the exciting x-ray field. Any deviation from this equality therefore indicates excitation beyond the \ler.

Next, we derive the correspondence between the population and coherence for effective nuclear two-level systems driven by general time-dependent x-ray fields. To this end, we expand Eqs.~(\ref{eq:SelfConsNoInit}) in powers of the driving x-ray field amplitude.

In the absence of the x-ray field, the nuclei are in their ground state, 
\begin{align}
    \rho_{ee}^{(0)}(t=t_0) &= 0\,, \qquad  \rho_{ge}^{(0)}(t=t_0) = 0\,, \nonumber\\
    \rho_{eg}^{(0)}(t=t_0) &= 0\,, \qquad \rho_{gg}^{(0)}(t=t_0) = 1\,.  \nonumber
\end{align}
Iteratively solving the self-consistent equations in a perturbative expansion, we find in first order in the driving x-ray field that (see Appendix~\ref{app:SelfDeriveSol})
\begin{subequations}
\begin{align}
\rho^{(1)}_{ge}(t) &= \int_{t_0}^t d\tau \: f(t, \tau)\,, \label{rho-eg-1}
 \\
\rho^{(1)}_{ee}(t) &= 0\,,
\end{align}
\end{subequations}
where we defined the function $f(t, \tau)$ as the integrand in Eq.~(\ref{rho-eg-1}) as
\begin{align}
f(t, \tau) = -\frac{i}{2} e^{-\frac{\gamma}{2}(t-\tau)}e^{i\omega_0(t-\tau)}\: \Omega^*(\tau)\,,
\end{align}
for reasons which will become clear in the next step. Note that  Eq.~(\ref{rho-eg-1}) bears similarity to the lowest-order result in Sec.~\ref{sec:simple-example} involving the pulse area in Eq.~(\ref{eq:Area}) in that the off-diagonal density matrix element depends on an integral over the coupling Rabi frequency over time. However, the more complete analysis here allows us to incorporate more degrees of freedom in our analysis such as detunings between exciting x-ray pulse and two-level system or decay processes.

Going further in the perturbative expansion, we find that the second-order contributions can be written as (see Appendix~\ref{app:SelfDeriveSol}),
\begin{subequations}
\begin{align}
\rho^{(2)}_{ge}(t) &= 0\,,\\[2ex]
\rho^{(2)}_{ee}(t) &= \left\lvert \int^{t}_{t_0}\: f(t, \tau) \:d\tau \right\rvert^2 = \left\lvert\rho^{(1)}_{ge}(t)\right\rvert^2
\label{eq:NestedPop}\,.
\end{align}
\end{subequations}

Hence, we find that the excited-state population and the coherence-squared are identical up to second order in the driving x-ray field  also in this more general case, 
\begin{align}
\rho^{(0-2)}_{ee}(t) = \left\lvert \rho^{(0-2)}_{ge}(t)\right\rvert^2\,, \label{eq:PopCohId}
\end{align}
where the superscript $(0-2)$ indicates that all contributions up to second order are included. 

As a result, we have shown for general two-level systems including decay and driven by weak time-dependent x-ray pulses that the coherently emitted  intensity is equivalent to the intensity of the incoherent emission in second order of the x-ray-nucleus coupling. Note that this result is independent of the temporal shape of the x-ray field, such that it also holds for pulse sequences. 

In terms of experimentally-accessible quantities, this implies that the ratio of the coherently and incoherently scattered intensities Eqs.~(\ref{eq:Iinc}),(\ref{eq:Icoh}) is constant as function of time in the LER, i.e., if both observables are expanded up to second order in the driving x-ray field. 
Conversely, deviations from this time-independence therefore imply dynamics beyond the LER. In particular, as discussed in Sec.~\ref{sec:simple-example}, the ratio is expected to change quadratically with the integrated Rabi frequency in leading higher-excitation order. In Secs.~\ref{sec:impulsive} and \ref{sec:non-impulsive}, we will explore particular x-ray pulse examples of experimental relevance for such dynamics beyond the LER, which will also allow for analytical solutions of the nuclear dynamics to higher order.

\subsection{Perturbative solution of the interacting many-body nuclear ensemble \label{sec:PerturbInt}}

In Appendix~\ref{sec:ManyBodyProof} we show that the relevant solution to the full master equation Eq.~(\ref{eq:Master}) of the $N$-body system up to second order in the x-ray-nuclei interaction, which at initial time $t_0$ is in the ground state, can be written for nucleus $x$  as,
\begin{subequations}
\label{sol-nbody-maintext}
\begin{align}
\rho_{g_x,e_x}^{(1)}(t) &=   \sum_{n=1}^N \int^{t}_{t_0}d\tau \:  g_{nx}(\tau)\,,   \\[2ex]
g_{nx}(\tau) &=-\frac{i}{2} \Omega^*(\mathbf{r}_n,\tau) \left[ e^{\expmatrix(t-\tau)} \right]_{nx}\,, \\[2ex]
\expmatrix &= (\notheta_{nm}) \, , \\[2ex]
\notheta_{nm} &= -(\Gamma_{nm} + i\,J_{nm})  - (\Gamma_{\mathrm{IC}} - i \omega_0)\delta_{nm}\,,\\[2ex]
\rho_{e_x,e_x}^{(2)}(t) &= \left | \rho_{e_x, g_x}^{(1)}(t)\right|^2\,,
\end{align}
\end{subequations}
where $\expmatrix$ is a matrix with elements $\notheta_{nm}$. With this result at hand, we can now proceed by showing the equivalence between the population-based observables and the coherence-based observables also in the many-body case.

The solutions Eqs.~(\ref{sol-nbody-maintext}) directly prove the equivalence on the single-particle level,
\begin{subequations}
\label{eq:ManyPopCohId-main}
\begin{align} 
\lvert \rho^{(1)}_{g_xe_x}(t)\rvert^2 &= \rho^{(2)}_{e_xe_x}(t)\,, \\[2ex]
\Rightarrow \lvert \rho^{(0-2)}_{g_xe_x}(t)\rvert^2 &= \rho^{(0-2)}_{e_xe_x}(t)\,.
\end{align}
\end{subequations}

However, this result on the single-particle level is not sufficient for the experimentally accessible ensemble-based observables, given by Eqs.~(\ref{eq:Iinc}) and (\ref{eq:Icoh}) as
\begin{align}
I_{inc}(t) &\propto \sum_{n=1}^N \: \langle \hat \sigma^+_n \hat \sigma^-_n\rangle \,,\\
I_{coh}(t, \mathbf{k}_{out}) &\propto \left\lvert \sum_n e^{i\mathbf{k}_{out}\mathbf{r}_n}\langle \hat{\sigma}^+_{n}\rangle \right\rvert^2 \,.
\end{align}
To evaluate the sums in these expressions, we make two assumptions: First, we assume that all nuclei are excited  with the same amplitude   using a plane-wave field with wave vector $\mathbf{k}_{\mathrm{in}}$, 
\begin{align}
  \Omega(\mathbf{r}_n,t) = \Omega(t) \, e^{i\mathbf{k}_{\mathrm{in}}\,\mathbf{r}_n}\,.
\end{align}
Second, we make the assumption of a homogeneous ensemble of nuclei, i.e.,
\begin{align}
&\rho_{g_x,e_x}^{(1)}(t) =  -\frac{i}{2} \sum_{n=1}^N \int^{t}_{t_0}d\tau \:  \Omega^*(\mathbf{r}_n,\tau) \left[ e^{ \expmatrix(t-\tau)} \right]_{nx} \nonumber \\[2ex]
=&  -\frac{i}{2} e^{-i\mathbf{k}_{\mathrm{in}}\,\mathbf{r}_x} \sum_{n=1}^N \int^{t}_{t_0}d\tau \: \Omega^*(\tau) e^{i\mathbf{k}_{in}(\mathbf{r}_{x}-\mathbf{r}_{n})}  \left[ e^{ \expmatrix(t-\tau)} \right]_{nx} \nonumber \\[2ex]
=&  -\frac{i}{2} e^{-i\mathbf{k}_{\mathrm{in}}\,\mathbf{r}_x} \sum_{n=1}^N \int^{t}_{t_0}d\tau \: \Omega^*(\tau) e^{i\mathbf{k}_{in}(\mathbf{r}_{n_0}-\mathbf{r}_{n})}  \left[ e^{ \expmatrix(t-\tau)} \right]_{nn_0} \nonumber \\[2ex]
=& \rho_{ge}^{(1)}(t)\:  e^{-i\mathbf{k}_{\mathrm{in}}\,\mathbf{r}_x}\,.
\end{align}
Here, in the second step, we have replaced the index $x$ in the integrand with an (arbitrary) fixed index $n_0$. This homogeneity assumption requires that the coupling environment of all two-level systems is equivalent and that the nuclei themselves are identical. Note that even in a regular arrangement, nuclei at the boundary of the medium experience different couplings to other nuclei than those in the center of the ensemble. But in macroscopically large ensembles, these boundary effects can be neglected to a good approximation.  Within the same assumptions, we also find
\begin{align}
    \rho^{(2)}_{e_xe_x}(t) = \rho^{(2)}_{ee}(t)\,,
\end{align}
i.e., the nuclear populations evolve independent of the atom index $x$. 

As a result, the result up to second order in the light-matter coupling is
\begin{subequations}
\label{eq:ManyBodyObs}
\begin{align}
I_{inc}^{(0-2)}(t) &\propto N\: \rho^{(2)}_{ee}(t) \,,\\[2ex]
I_{coh}^{(0-2)}(t, \mathbf{k}_{out}) &\propto \left\lvert\rho_{ge}^{(1)}(t)\right\rvert^2\: \cdot \left\lvert \sum_{n=1}^N e^{i(\mathbf{k}_{out}-\mathbf{k}_{in})\mathbf{r}_n} \right\rvert^2 \,. \label{eq-geom}
\end{align}
\end{subequations}
Together with Eq.~(\ref{eq:ManyPopCohId-main}), we thus again find that in second order of the x-ray-nucleus coupling, the ratio of the coherently scattered x-rays to those incoherently scattered is constant in time,
\begin{align}
\frac{I_{coh}^{(0-2)}(t, \mathbf{k}_{out})}{I_{inc}^{(0-2)}(t)} = \frac{I_{coh}^{(0-2)}(t_0, \mathbf{k}_{out})}{I_{inc}^{(0-2)}(t_0)}\,.
\end{align}
Note that the geometrical factor in Eq.~(\ref{eq-geom}) characterizes the directionality of the coherently scattered light. 

\section{\label{sec:impulsive}Impulsive x-ray excitation beyond the low-excitation regime}

\subsection{Two-level analysis~\label{two-level-impulsive}}
In the impulsive excitation case, the x-ray excitation and the subsequent decay dynamics in the absence of a driving x-ray field can be considered separately. For a single particle, the initial state after the x-ray excitation is given by  Eqs.~(\ref{eq:PulseAreaNoInit})  as  
\begin{subequations}
\begin{align} 
\rho_{ee}(t=0) &= \sin^2\left[\mathcal{A}\right] \,,  \\
\rho_{ge}(t=0) &= -\frac{i}{2}\, e^{i\phi}\,\sin\left[2\mathcal{A}\right]\,.
\end{align}
\end{subequations}
Here, for simplicity, we denote the time after the exciting x-rays have passed the nuclei as $t=0$, and $\mathcal{A}$ is the total pulse area of the exciting x-ray pulse. Subsequently, the decay is governed by~\cite{Scully1997}

\begin{subequations}
\begin{align}
\rho_{ee}(t) &= \rho_{ee}(0)\, e^{-\gamma t}\,,  \\
\rho_{ge}(t) &= \rho_{ge}(0)\, e^{-\frac{\gamma}{2} t}\,.
\end{align}
\end{subequations}
As a result, we find that the ratio of the coherent and incoherent intensities Eqs.~(\ref{eq:Icoh}) and (\ref{eq:Iinc}) evaluates to
\begin{align}
\frac{I_{coh}(t)}{I_{inc}(t)}\propto \frac{\left  |\rho_{ge}(t)\right|^2}{\rho_{ee}(t)} = \frac{\left  |\rho_{ge}(0)\right|^2}{\rho_{ee}(0)} = \cos^2\left[\mathcal{A}\right]\,.
\end{align}
Hence, the ratio remains constant over time, but depends on the degree of excitation (cf. Fig.~\ref{fig:CoupledRatios}). Measuring it as a function of the resonant intensity of the exciting x-rays then allows one to search for deviations in this ratio from its value at low x-ray intensities, which indicate excitations beyond the \ler. One possibility for this measurement is to exploit the typically large pulse-to-pulse fluctuations in resonant intensity at x-ray free electron lasers, and to sort the intensity data according to the incident (resonant) pulse energy.

\subsection{\label{sec:CoupDyn}Dynamics of a coupled nuclear ensemble after impulsive excitation}

If the interactions between the nuclei are weak enough such that the dynamics of the individual nuclei is essentially independent of each other on nuclear decay timescales, then the results of the single two-level nucleus case are recovered for each nucleus separately. 

A more interesting situation arises in case of stronger couplings. For example, a suitably designed~\cite{haber_collective_2016,Haber2017,PhysRevA.106.053701,PhysRevA.105.013715} waveguide structure may allow one to realize regimes of stronger interactions between the nuclei. 

In Sec.~\ref{sec:PerturbInt} we showed that in this regime, the coherence-population correspondence Eq.~(\ref{eq:PopCohId}) is still valid up to second order in the nucleus-field interaction. This holds  for each nucleus separately and, more importantly, also the scaled ratio of the total incoherently and coherently scattered intensity 
\begin{align}
\mathcal{R}(t) = \frac{1}{N}\frac{\lvert \sum_{n}e^{i\mathbf{k}_{out}\mathbf{r}_n} \rho_{g_ne_n}(t)\rvert^2}{\sum^N_{n=1} \rho_{e_ne_n}(t)} \label{eq:r-scaled}
\end{align}
is constant in time and equal to one if the two assumptions of excitation by a plane wave field and a homogeneous nuclear ensemble are satisfied (see Sec.~\ref{sec:PerturbInt}). Note that  $\mathcal{R}(t)$ is proportional to the intensity ratio $R(t)$ (see Fig.~\ref{fig:main}), and  scaled  by the number of nuclei $N$ since the coherently scattered intensity is proportional to $N^2$ in the ideal case of full constructive interference, while the incoherent part is proportional to $N$. 

\begin{figure}[t]
\includegraphics[width = 0.95\columnwidth]{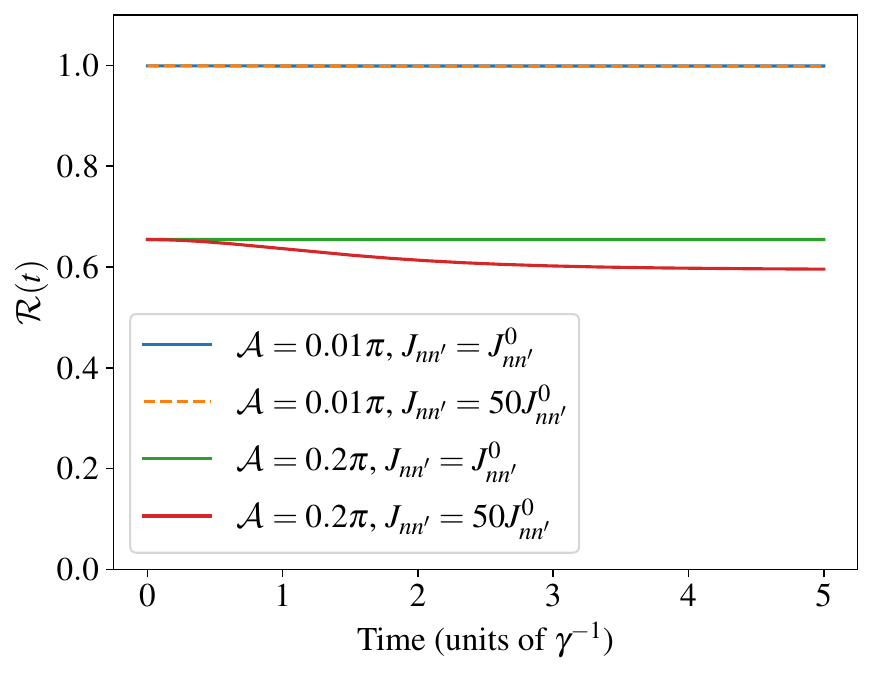}
\caption{Ratio $\mathcal{R}(t)$ of the coherently and incoherently scattered intensity for a regular chain of $N=8$ nuclei with periodic boundary conditions, scaled by $N$ [see Eq.~(\ref{eq:r-scaled})]. The lattice constant $r_0 = 286$~pm and the resonant wavelength $\lambda_0 = 86$~pm are chosen for the case of $\alpha$-iron enriched in ${}^{57}$Fe to determine the  coherent dipole-dipole coupling constants $J^0_{nn'}$. Results are compared for the \ler ($\mathcal{A} = 0.01\pi$) and stronger excitation ($\mathcal{A} = 0.2\pi$), as well as for free-space coupling ($J_{nn'}^0$) and cavity-enhanced coupling ($50\,J_{nn'}^0$). Note that the two curves in the \ler case coincide, as expected. \label{fig:CoupledRatios}}
\end{figure}

Next, we analyze the dynamics beyond the \ler, also in the presence of stronger couplings between the nuclei, using a numerical integration of the full master equation for a limited number of nuclei. The calculations are performed using the \textsc{python} library QuTiP~\cite{JOHANSSON20121760,JOHANSSON20131234}. 
In particular, we consider the dynamics of a linear chain of $N=8$ regularly arranged nuclei with periodic boundary conditions after impulsive x-ray excitation. The coherent dipole-dipole coupling parameters $J_{nn'}$ are chosen assuming nuclear dipole moments $\hat{\mathbf{d}}$ oriented perpendicularly to the relative positions $\mathbf{r}_n-\mathbf{r}_{n'}$ of the nuclei in the chain. Their free space values can then be calculated via~\cite{Agarwal1974,Ficek_Swain}
\begin{align}
J^0_{nn'} = \frac{3}{2} \Gamma \left(\frac{\cos(\eta_{nn'})}{\eta_{nn'}}-\frac{\sin(\eta_{nn'})}{\eta_{nn'}^2}-\frac{\cos(\eta_{nn'})}{\eta_{nn'}^3}\right)\,,
\end{align}
where $\Gamma = \Gamma_{nn} = \gamma/[2(1+\alpha)]$ denotes the radiative decay rate of the nuclei. Here, $\alpha = 8.56$ is chosen as the internal conversion coefficient of the archetype M{\"o}ssbauer isotope ${}^{57}$Fe with the  M\"ossbauer resonance at 14.4~keV transition energy~\cite{RoehlsbergerBook}. We further defined $\eta_{nn'} = k_0\,r_{nn'}$. The resonant wave number $k_0 = 2\pi/\lambda_0$ and the distances $r_{nn'} = \lvert n-n'\rvert\,r_0$ between the nuclei are chosen corresponding to the resonant wavelength $\lambda_0 = 86$~pm and the lattice constant $r_0 = 286$~pm for $\alpha$-iron enriched in ${}^{57}\textrm{Fe}$.
Note that for a small ensemble with periodic boundary conditions, the homogeneity criterion for the initial phases imprinted on the nuclei by the x-ray excitation can only be satisfied for particular incidence angles. To satisfy this criterion, we chose the incidence angle such that a relative phase of $k_0\,r_{n, n+1} = 2\pi/N$ is imprinted onto the coherence of neighbouring nuclei. For the same reason the decay rates $\Gamma$ and nuclear transition frequencies $\omega_0$ are considered to be the same for all nuclei in the chain. Further, incoherent couplings $\Gamma_{nn'}$ were neglected in the numerical simulation.

Figure~\ref{fig:CoupledRatios} shows the simulation results for different dipole-dipole coupling parameters $J_{nn'}$ and pulse areas $\mathcal{A}$. The chosen parameters correspond to the \ler case ($\mathcal{A} = 0.01\pi$) or excitation beyond the \ler ($\mathcal{A} = 0.2\pi$), as well as weak dipole-dipole coupling ($J_{nn'} = J_{nn'}^0$) and coupling enhanced by a factor of 50 relative to the free-space coupling ($J_{nn'} = 50\,J_{nn'}^0$).

In the \ler case (blue and orange-dashed lines), we find that the ratio $\mathcal{R}(t)=1$, independent of the dipole-dipole coupling, consistent with our analytical results in Sec.~\ref{sec:PerturbInt}. 
For excitation beyond the \ler case in the presence of weak coupling (green curve), we find a constant ratio $\mathcal{R}$, however, with a value below 1. This agrees with our results in Sec.~\ref{two-level-impulsive}. 
In contrast, for excitations beyond the \ler with stronger dipole-dipole couplings (red curve), the ratio $\mathcal{R}(t)$ becomes time-dependent and evolves to lower values, initially starting from the ratio for the low-coupling case. This is due to a faster transient decay of the coherences entering the expression for the coherently scattered intensity in the interacting system. 

As a result, we conclude that the time-dependent ratio of the coherently- and incoherently scattered x-rays does not only serve as a criterion for the excitation beyond the \ler, but may further also reveal the presence of stronger dipole-dipole couplings between the nuclei.

\section{\label{sec:non-impulsive}Non-impulsive x-ray excitation beyond the low-excitation regime}
 
For the case of impulsive excitation, we found that a comparison of the coherently and incoherently scattered intensity provides a handle to identify excitation of the nuclear ensemble beyond the \ler. In case of effective single-particle dynamics, the ratio between these two intensities remains constant throughout the decay while it becomes time-dependent at higher excitations in sufficiently-strongly interacting nuclear ensembles. In the following, we extend this discussion to the non-impulsive regime. The calculation of the complete dynamics of a large ensemble of coupled nuclei under the action of time-dependent driving fields and dissipation so far is an unsolved problem, and remains beyond the scope of this work. Instead, we analyze the non-impulsive dynamics in the (effective) single-particle case. It is expected that the single-particle description, which is valid in the LER~\cite{lentrodt_ab_2020}, remains a good approximation at the onset of dynamics beyond the \ler but likely breaks down at higher excitation. Nevertheless, in the following, we also explore the dynamics at stronger excitation using the single-particle description, with the motivation of identifying possible experimental signatures for dynamics beyond the \ler.  Using this approach, the results presented in this Section are then obtained by numerically integrating the optical Bloch equations Eqs.~(\ref{eq:Bloch}).

To analyze the non-impulsive case, we consider nucleus-field couplings of the form
\begin{align} \label{eq:MonoPulse}
\Omega(t) = 2\,\Gamma_a\, \mathcal{A}\:e^{-i(\omega_0+\Delta)t}\:e^{-\Gamma_at}\,,
\end{align} 
which characterizes an x-ray pulse exponentially decaying with rate $\Gamma_a$ and center frequency detuned by $\Delta$ from the nuclear resonance frequency $\omega_0$. Its total pulse area according to Eq.~(\ref{eq:Area}) is given by $\mathcal{A}$.

\begin{figure}[t]
\includegraphics[width = 0.95\columnwidth]{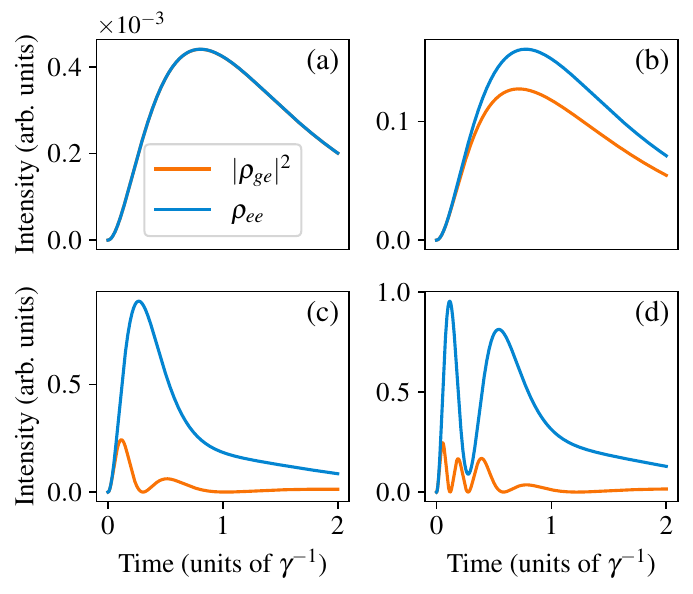}
\caption{Temporal dynamics of the excited-state population $\rho_{ee}$ and the coherence squared $|\rho_{ge}|^2$ of an effective two-level system for different degrees of excitation. The population and the coherence squared relate to the observable incoherently and coherently scattered light intensities.  The different panels correspond to driving x-ray pulse areas of (a) $\mathcal{A} = 0.01\pi$, (b) $\mathcal{A} = 0.2\pi$, (c) $\mathcal{A} = \pi$, and (d) $\mathcal{A} = 2\pi$. 
In the  \ler [panel (a)], population and coherence squared agree, consistent with our analytical results. Upon excitation beyond the \ler, the two quantities start to deviate (b), and eventually Rabi oscillations appear (c,d). Note that in (d), the oscillation frequencies of the coherence squared and of the population differ approximately by a factor of 2. \label{fig:RabiApp}}
\end{figure}

This choice for the driving x-ray field is motivated by the availability of synchrotron M\"ossbauer sources (SMS)~\cite{gerdau_nuclear_1985,PhysRevB.55.5811,PhysRevLett.102.217602,masuda_development_2008,potapkin_57fe_2012} which employ pure nuclear reflexes to produce x-ray pulses which are spectrally narrow on nuclear line-width scales from the incident broadband synchrotron pulses. In the future, these sources could be generalized for operation at x-ray free electron lasers~\cite{chumakov_superradiance_2018}. Another source providing spectrally narrow pulse contributions is the field scattered in forward direction by thin nuclear targets in the LER, which is approximately exponentially-decaying~\cite{Smirnov1986,Hannon1999}. By moving the thin nuclear target before or throughout its decay, the properties of the scattered light relative to those of the incident synchrotron pulse can be tuned~\cite{Helistoe1991,PhysRevB.47.7840,Schindelmann2002,Vagizov2014,Heeg2017,Heeg2021,Shvydko2022,Heeg2022,PhysRevLett.123.250504,Liao2017,PhysRevLett.122.025301}. In particular, using suitably tailored x-ray pulses, the quantum dynamics of a nuclear target could be controlled~\cite{Heeg2021}. A possible generalization of such schemes to higher excitation at x-ray free electron lasers again requires further analysis of nuclear dynamics under exponentially-decaying x-ray pulses such as in Eq.~(\ref{eq:MonoPulse}). 

As expected, we will find that the non-instantaneous driving field gives rise to a much richer dynamics than in the impulsive case, since the x-ray-induced  dynamics and the decay dynamics are not temporally separated in the former case. In particular, this will affect the time-dependence of the ratio of the coherently and incoherently scattered intensities.

Note that this time-dependence of the effective single-particle observables will also be reflected in the corresponding many-body observables as can be seen from Eqs.~(\ref{eq:ManyBodyObs}) such that their ratio will become time-dependent as soon as the single-particle quantities are.

\begin{figure}[t]
\includegraphics[width = 0.95\columnwidth]{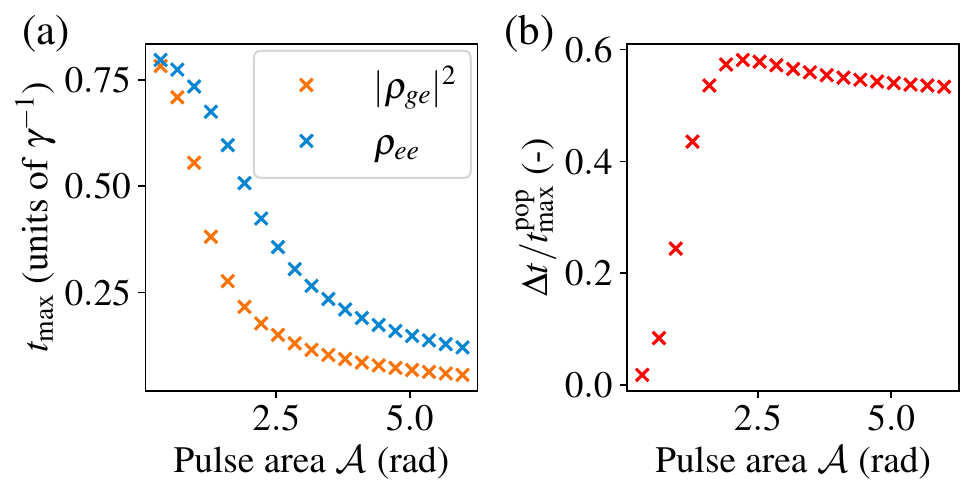}
\caption{Analysis of the times with maximum intensities in the coherence squared and the populations. (a) shows the times defined in Eqs.~(\ref{eqs:MaxTimes}) as function of the total pulse area $\mathcal{A}$. The other parameters are as in Fig.~\ref{fig:RabiApp}. While the maxima in the population and the coherence squared coincide in the \ler, deviations appear towards stronger excitation. (b) shows the peak deviation defined in Eq.~(\ref{eq:PeakDev}) against $\mathcal{A}$. Starting from zero value in the \ler, the deviation steeply increases at the onset of non-linear excitation of the nuclear ensemble. \label{fig:PeakDiff}}
\end{figure}

\subsection{Resonant case $\Delta = 0$}

We start by analyzing the case of a resonant x-ray pulse Eq.~(\ref{eq:MonoPulse}) with detuning $\Delta=0$ and decay constant $\Gamma_a = 2.5 \gamma$, where $\gamma$ denotes the total line-width of the effective single nucleus. Results for different pulse areas $\mathcal{A} \in \{ 0.01 \pi, 0.2\pi, \pi, 2\pi\}$  are shown in Fig.~\ref{fig:RabiApp}. 

Consistent with the analytical results in Sec.~\ref{sec:ProofCorrespond}, the coherence squared and the population characterizing the  coherently and  incoherently scattered intensity, respectively, agree in the low-excitation case [panel (a), $\mathcal{A} = 0.01 \pi$]. With increasing excitation, deviations between the two observables start to appear [see panel (b)]. 
If the driving field becomes strong enough to induce Rabi oscillations [panels (c) and (d)], the different oscillation periods of coherence quared and population become visible. As a result, time-dependent ratios of the coherently and incoherently scattered light intensities can be expected in the non-impulsive driving case beyond the \ler. 

From these results we find that a first signature for dynamics slightly beyond the \ler is a relative shift in the peak maxima of the two time-dependent intensities. This relative shift arises from the competition of the coherent excitation dynamics and the incoherent decay dynamics.  To analyze this shift more quantitatively, we define the two corresponding times $t^{\textrm{coh}}_{\textrm{max}}$ and $t^{\textrm{pop}}_{\textrm{max}}$  with maximum intensities via the conditions
\begin{subequations}\label{eqs:MaxTimes}
\begin{align}
\lvert \rho_{ge}(t^{\textrm{coh}}_{\textrm{max}}) \rvert^2 =& \max_{t \in [0,\infty)}(\lvert\rho_{ge} (t)\rvert^2)\,, \\[2ex]
\rho_{ee}(t^{\textrm{pop}}_{\textrm{max}}) =& \max_{t \in [0, \infty)}(\rho_{ee}(t))\,.
\end{align}
\end{subequations}
We further consider a peak deviation defined as
\begin{align} \label{eq:PeakDev}
\Delta t / t^{\textrm{pop}}_{\textrm{max}} =  \frac{\lvert t^{\textrm{coh}}_{\textrm{max}} - t^{\textrm{pop}}_{\textrm{max}} \rvert }{t^{\textrm{pop}}_{\textrm{max}} }\,.
\end{align}
Note that multiple maxima may appear due to Rabi oscillations at stronger x-ray driving. In this case, we consider the respective maxima appearing first after the onset of excitation. 

Figure~\ref{fig:PeakDiff} shows the times $t^{\textrm{coh}}_{\textrm{max}}$ and $t^{\textrm{pop}}_{\textrm{max}}$ [panel~(a)] and the corresponding peak deviation [panel~(b)] as a function of the total pulse area $\mathcal{A}$. The deviations in case of dynamics beyond the \ler are clearly visible. In particular, the steep increase of the peak deviation at small pulse areas renders it a promising signature for characterizing dynamics at the onset of the nonlinear excitation regime.

\begin{figure}[t]
\includegraphics[width = 0.95\columnwidth]{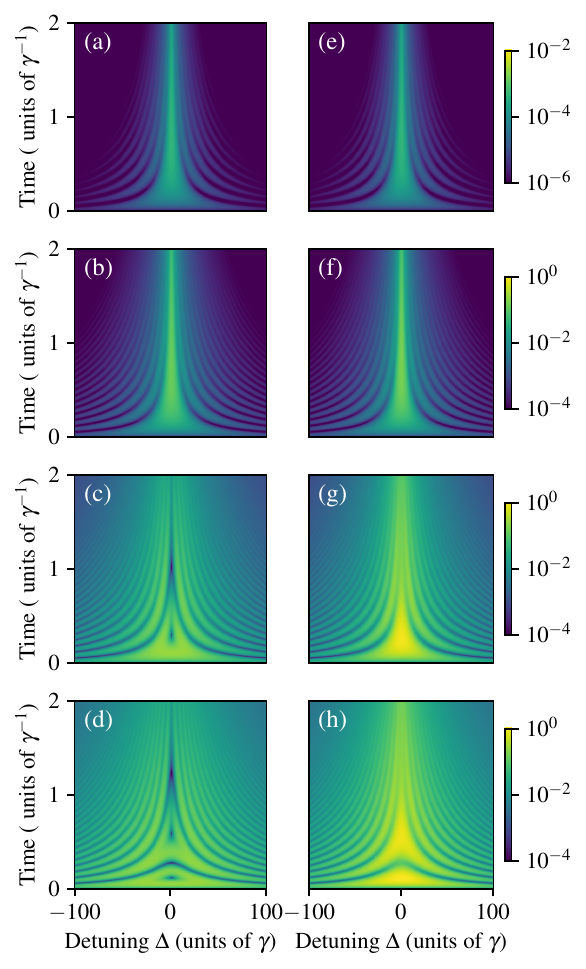}
\caption{Time- and frequency-resolved coherence squared (left panels (a) - (d)) and populations (right panels (e) - (h)) for the case of non-impulsive excitation with an exponentially decaying pulse Eq.~(\ref{eq:MonoPulse}). The pulse envelope decays with $\Gamma_a = 2.5\gamma$. The rows from top to bottom are calculated for pulse areas of $\mathcal{A} \in \{0.01\pi, 0.2\pi, \pi, 2\pi\}$. \label{fig:2dcomparison}}
\end{figure}

\subsection{Non-resonant case $\Delta \neq 0$ \label{sec:Nonres}}

Next, we generalize to the non-resonant case $\Delta \neq 0$. To this end, we analyze the coherence squared $|\rho_{ge}|^2$ and the population $\rho_{ee}$ as a function of time and detuning $\Delta$. This correlated analysis of temporal- and spectral properties has proven to be a powerful tool in analyzing nuclear resonant scattering~\cite{Heeg2017,Heeg2021,PhysRevResearch.5.013071}. However, it is important to note that the experimental setup underlying the theoretical analysis of this manuscript is different from previous approaches to record such time- and frequency-resolved spectra. 
Previous studies  considered an \textit{impulsive} x-ray excitation of the nuclear ensemble. The frequency resolution in these setups is achieved via an additional frequency-tunable reference absorber with an approximately exponential temporal decay. As a result, the time-frequency spectra are dominated by the interference of the light scattered by the reference absorber and the target, respectively~\cite{PhysRevResearch.5.013071}. 

In contrast, here, we consider time-frequency spectra for the target driven by a \textit{non-impulsive} frequency-tunable driving pulse of the form Eq.~(\ref{eq:MonoPulse}), in the absence of additional reference absorbers.  This setup could be realized, e.g., using a synchrotron-M\"ossbauer-source, by correlating the detuning $\Delta$ of the source with the time-dependence of the scattered photons.

Figure~\ref{fig:2dcomparison} compares the time-frequency spectra for different pulse areas $\mathcal{A}$. Again, a pulse decay rate of $\Gamma_a = 2.5 \gamma$ is chosen in the calculation. The left panels (a-d) show the coherence squared, whereas the right panels (e-h) depict the population. In the \ler, the two signatures agree, as expected [top row, panels (a,e)]. Differences between the spectra appearing at larger pulse areas [panels (c,g) or (d,h)] are most pronounced at small detunings, which can be understood by noting that the excitation of the nuclei for a given pulse area is highest towards resonance, such that the deviations from the LER are more pronounced while higher excitation orders are suppressed further off-resonance. The visible deviations are a consequence of the Rabi oscillations discussed in the previous Section (cf. Fig.~\ref{fig:RabiApp}). Therefore, we conclude that also the energy-time correlation spectra may be used to identify the presence of excitation beyond the \ler. However, it turns out that these spectra contain additional signatures for a non-linear excitation, which can be revealed using a Fourier transform along the time axis, as discussed next.

\begin{figure}[t]
\includegraphics[width = 0.95\columnwidth]{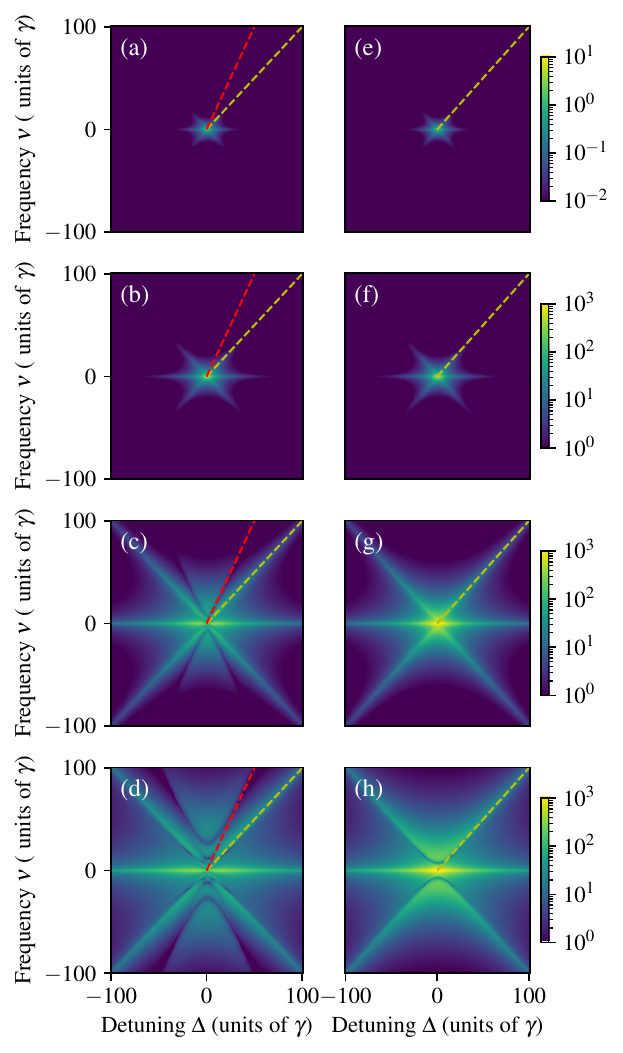}
\caption{ Frequency-frequency correlation spectra. The results are obtained via Fourier transforms along the time axes of the time-frequency spectra in Fig.~\ref{fig:2dcomparison}. 
Data is shown in color-coded logarithmic scale for the coherence squared [left panels, (a)-(d)] and for the populations [right panels, (e)-(h)]. As in Fig.~\ref{fig:2dcomparison}, the rows from top to bottom refer to different pulse areas of $\mathcal{A} \in \{0.01\pi, 0.2\pi, \pi, 2\pi\}$. The yellow [red] lines indicate slope one [two] as guide to the eye to facilitate the interpretation. \label{fig:FouLinComparison}}
\end{figure}

\subsection{Frequency-frequency correlation-spectra analysis of the non-resonant case $\Delta \neq 0$ \label{sec:FFC}}

The interpretation of the time- and frequency-resolved spectra in Fig.~\ref{fig:2dcomparison} is facilitated by a Fourier transform along the time axis, which yields a frequency-frequency correlation (\ffc) spectrum~\cite{PhysRevResearch.5.013071}. In particular, the hyperbolic structures in the time-frequency spectra convert into diagonal lines in the \ffc spectra, which greatly assists their analysis.  However, we stress again that the present manuscript considers {\it non-impulsive} x-ray driving pulses without additional reference absorbers, such that the results cannot directly be compared to previous studies (see the discussion in Sec.~\ref{sec:Nonres}).

Results for the \ffc spectra are shown in Fig.~\ref{fig:FouLinComparison}. For lower pulse areas, the diagonal structures in the FFC spectra are suppressed as compared to the impulsive case~\cite{PhysRevResearch.5.013071}, since the nuclear excitation at off-resonant energies is negligible, which is in contrast to the impulsive case, where resonant contributions from the reference absorber give rise to visible contributions at all detunings.
Towards higher pulse areas and stronger excitation, we recognize two distinct features. First, on resonance $\Delta =0$, the population and the coherence squared show qualitatively different behavior, because of their different Rabi oscillation frequencies in the time domain (cf. Sec.~\ref{sec:simple-example}). The appearance of these differences again serves as a signature for dynamics beyond the \ler. 
Second, off-resonance ($\Delta \neq 0$), diagonal structures start to appear. Interestingly, the population only exhibit diagonals of slope one (yellow lines), corresponding to a dependence of its dynamics on the detuning~\cite{PhysRevResearch.5.013071}. In contrast, the coherence squared exhibits diagonals with slope one (yellow lines) and two (red lines), corresponding to contributions oscillating with frequencies $\Delta$ and $2\Delta$.
The additional pair of diagonals with slope two arises from a frequency-mixing of different scattering orders in squaring the coherence. In Appendix~\ref{sec:NonresAna}, we analyze this feature further, and show that in the \ler, only the diagonals of slope one appear in the coherence squared. Starting from the next higher order, diagonals of both slopes become visible. Therefore, the appearance of diagonals with slope two are a clear qualitative signature for excitation beyond the \ler. 
At even higher pulse areas, the Rabi oscillations dominate, and the diagonals in the FFC spectra develop an anti-crossing-like feature towards their center, which can be attributed to the onset of an Autler-Townes splitting of the resonance.

\section{Discussion and summary\label{sec:summary}}

In this work we have developed methods to verify the excitation of an ensemble of M\"ossbauer nuclei beyond the low-excitation regime (\ler). This is motivated by the recent availability of x-ray free-electron lasers which are capable of delivering many resonant photons per pulse, such that the as-yet unexplored regime of stronger excitation of the nuclei comes within reach. Since source limitations will likely persist at least in the near future, we focused our analysis on  practically relevant and experimentally robust approaches which are applicable already in a regime where the linear x-ray-nuclei interaction is only slightly surpassed. 

Our approach is based on the comparison of two observables, which can be measured concurrently in an experiment: the coherently scattered light, which features a directional emission in a narrow angular range, and the incoherently emitted internal conversion signatures (photons or electrons) which are emitted essentially into the full solid angle. The latter signature is proportional to the sum of the excited-state populations of the nuclei, while the former is related to the absolute value squared of the coherent sum of the induced dipole moments augmented by phase factors related to the excitation and de-excitation of the nuclei. 

As a key step, we established that the time-dependent intensities emitted into the coherent and the incoherent channels are equivalent up to second-order in the driving x-ray field for a wide range of systems, such that their ratio is constant in time. To this end, we first proved the corresponding equivalence between the populations and coherences for an effective single-particle system describing a nuclear ensemble embedded in a waveguide environment. Subsequently, we extended the proof to a general  homogeneous $N$-body system of interacting nuclei. As a result, we found that  any deviations in the ratio between the x-rays coherently and incoherently scattered off of the nuclei can directly be traced back to an excitation beyond the \ler.  

Throughout our analysis, we considered two relevant cases: First, a  near-instantaneous impulsive excitation, e.g., via an XFEL. Second, a more general non-impulsive excitation, such as an exponentially decaying pulse from a synchrotron-M\"ossbauer-like source. The general results for the coherence-population equivalence in second-order of the driving x-ray field hold in both cases. We further focused on the case of nuclei embedded in waveguides, for which the observables are largely unaffected by propagational effects. It remains an interesting open question whether a generalized  equivalence between coherently- and incoherently scattered intensity can also be established in the presence of propagational effects.

We found that the various considered settings feature different dynamics in case of excitation beyond the \ler. For an impulsively excited single effective level scheme, the ratio between coherently and incoherently scattered intensity changes with the degree of excitation, but it remains constant over time. Analogously, the ratio also depends on the degree of initial impulsive excitation in the interacting many-body system. However, the many-body system further exhibits a time-dependence of the intensity ratio if the excitation exceeds the \ler and if simultaneously the inter-particle coupling is sufficiently enhanced beyond the free-space value, e.g., via a suitable waveguide environment. 

In case of non-impulsive excitation, we found that the time-dependence of the intensity ratio becomes much richer. To this end, we numerically studied the temporal dynamics of a single effective level scheme driven by an exponentially-decaying x-ray pulse. Based on the results, we proposed several signatures which can be used to identify excitation beyond the \ler. In the case of a resonant driving field, the first indication are the respective times at which the two intensities become maximal. These times agree in case of low excitation, but characteristic deviations appear for stronger excitation. At even higher excitation Rabi oscillations typical for the strongly nonlinear regime dominate the dynamics of both scattering observables, however, with different respective oscillation frequencies.  We further studied off-resonant excitation, using time-frequency correlated and frequency-frequency correlated spectra. In the time-frequency spectra, near resonance again resonant Rabi oscillations appear as the strongest indicator of excitation beyond the \ler, while the off-resonant regime strongly resembles the \ler even at higher excitations. In the frequency-frequency correlation spectra, characteristic diagonal structures appear. We could show that the incoherent intensity only exhibits a single pair of diagonals, with slope one. In contrast, at excitation beyond the \ler, the coherently-scattered intensity shows two pairs of diagonals at slopes one and two. Thus, the characteristic diagonal structure may also serve as a signature for stronger excitation of the nuclear ensemble.

Overall, we therefore conclude that the ratio of the coherently and incoherently scattered intensities serves as a strong indicator for excitation of a nuclear ensemble beyond the \ler. Its time structure can further reveal the presence of dipole couplings between nuclei.

In general, our approach has the advantage that both observables, the coherently and the incoherently scattered light, are well-established in experiments with M\"ossbauer nuclei. Most related experiments at accelerator-based x-ray sources have focused on the coherently scattered intensity, but the incoherently scattered intensity typically could be measured in addition without changing the original setup significantly. This suggests the usefulness of our approach even for  experiments in which the degree of excitation is not the primary research goal. A comparable situation existed in the traditional distinction between M\"ossbauer experiments measuring either in the time or in the energy domain. It was recently shown~\cite{Heeg2017,Heeg2021,PhysRevResearch.5.013071} that combined spectra correlating temporal and spectral information of the coherently scattered x-rays provide significant advantages over the individual time or frequency spectra, in particular also related to the comparison between theory and experiment. Also in that case, the original experimental setup could remain largely unchanged, and only had to be augmented by an event-based detection electronics. We envision that similar progress will be achieved in future experiments by the additional correlation of the coherently and incoherently scattered radiation proposed in the present work. 

\appendix

\section{Derivation of self-consistent equations for the coherence and population of a single two-level system \label{app:SelfDerive}}

In this appendix, we derive the self-consistent equations Eqs.~(\ref{eq:SelfConsNoInit}). Our approach is based on similar self-consistent approaches in quantum optics~\cite{Scully1997,MeystreSargent2007} and nonlinear optics~\cite{boyd2008nonlinear,Hamm2011}.

For a single two-level atom, the optical Bloch equations can be derived from the general $N$-body equations of motion Eq.~(\ref{eq:Master}) as
\begin{subequations}
\label{eq:Bloch}
\begin{align}
    \dot{\rho}_{ee}&=-\gamma \rho_{ee}+\frac{i}{2}\left[\Omega(t)\rho_{ge}-\Omega^*(t)\rho_{eg}\right]\,,  \label{ExEq}\\[2mm]
    \dot{\rho}_{ge}&=\big(i\omega_0-\frac{\gamma}{2}\big)\rho_{ge}+\frac{i\Omega^*(t)}{2}\left (2\rho_{ee}-1\right)\,, \label{CohEq} \\[2mm]
    \rho_{eg}&=\rho^{*}_{ge} \, , \quad\quad \rho_{gg}=1-\rho_{ee}\,,
\end{align}
\end{subequations}
where we have dropped the atom indices for notational simplicity, and $\gamma$ is the total natural line-width. Note that the same equations of motion govern the dynamics of effective two-level systems in the low-excitation sector of nuclear resonant scattering, however, with collectively modified parameters $\omega_0$, $\gamma$ and $\Omega(t)$.

To solve these equations self-consistently, we start by a formal integration. To this end, we introduce new variables
\begin{subequations}
    \begin{align}
\tilde{\rho}_{ge}(t) &= e^{(-i\omega_0 + \frac{\gamma}{2})t}\rho_{ge}(t) \,, \\
\tilde{\rho}_{ee}(t) &= e^{\gamma t} \rho_{ee}(t)\,.
\end{align}
\end{subequations}
Their time evolution is described by
\begin{subequations}
\begin{align}
\dot{\tilde{\rho}}_{ge}(t) &= \frac{i}{2}\:\Omega^*(t)\:\left [2\rho_{ee}(t)-1 \right]\,e^{-i\omega_0 t+\frac{\gamma}{2} t} \,,\\[1ex]
\dot{\tilde{\rho}}_{ee}(t) &= -\textrm{Im}\left[\Omega(t) \rho_{ge}(t)\right]\: e^{\gamma t}\,.
\end{align}
\end{subequations}
An integration from an initial time $t_0$ on yields
\begin{subequations}
\begin{align}
\tilde{\rho}_{ge}(t)-\tilde{\rho}_{ge}(t_0) &= \frac{i}{2}\int^{t}_{t_0} dt' \: \Omega^*(t')\: \left[2\rho_{ee}(t')-1\right] \nonumber \\
&\quad \qquad \times e^{-i\omega_0 t'+\frac{\gamma}{2} t'}\,,\\[1ex]
\tilde{\rho}_{ee}(t)-\tilde{\rho}_{ee}(t_0) =& -\int^{t}_{t_0}dt' \: \textrm{Im}\left[\Omega(t') \rho_{ge}(t')\right]\, e^{\gamma t'}\,.
\end{align}
\end{subequations}
Rewriting these equations in terms of the untransformed density matrix elements gives
\begin{subequations}
\begin{align}
    \rho_{ee}(t,t_0) &= e^{-\gamma (t-t_0)}\: \rho_{ee}(t_0) \nonumber \\
    &\quad -\int^t_{t_0}dt' \: e^{\gamma (t'-t)}\: \textrm{Im}\big[\Omega(t')\rho_{ge}(t',t_0)\big]\,, \\[2mm]
    \rho_{ge}(t,t_0) &= e^{(i\omega_0 - \frac{\gamma}{2}) (t-t_0)}\rho_{ge}(t_0) +\frac{i}{2}\int^t_{t_0}dt' \: \Omega^*(t') \nonumber \\[2mm]
    &\qquad  \times e^{-(i\omega_0-\frac{\gamma}{2}) (t'-t)}\,\left [2\rho_{ee}(t',t_0)-1\right ]\,,
\end{align}
\end{subequations}
where the initial density matrix element is defined as $\rho_{ij}(t_0):=\rho_{ij}(t_0,t_0)$. Inserting the second equation into the first one and vice versa yields:
\begin{subequations}
\label{FormalBoth}
\begin{align}
 \rho_{ee}&(t,t_0) = e^{-\gamma t}\left\{e^{\gamma t_0}\rho_{ee}(t_0)+e^{\frac{\gamma}{2} t_0}\int^{t}_{t_0}dt'e^{\frac{\gamma}{2} t'} \right. \nonumber \\[1ex]
 &  \times \mathrm{Re}\left[ie^{i\omega_0 (t'-t_0)}\Omega(t')\rho_{ge}(t_0)\right] \nonumber \\[1ex]
 & -\mathrm{Re}\Big[\int^{t}_{t_0}dt' \: e^{\gamma t'}\Omega^{*}(t')\int^{t'}_{t_0}dt'' \: e^{(i\omega_0+\frac{\gamma}{2}) (t''-t')} \nonumber \\[1ex]
 & \left. \left. \quad \times \Omega(t'')\left(\rho_{ee}(t'',t_0)-\frac{1}{2}\right)\right]\right\} \,, \label{FormalEx}\\[2ex]
 \rho_{ge}&(t,t_0)= e^{i\omega_0 t}e^{-\frac{\gamma}{2} t}\left\{e^{-i\omega_0 t_0}e^{\frac{\gamma}{2} t_0}\rho_{ge}(t_0) \right. \nonumber \\[2ex]
 &+i\int^{t}_{t_0}dt'e^{-i\omega_0 t'}\Omega^*(t')
    \big( e^{\gamma (t_0-\frac{t'}{2})}\rho_{ee}(t_0)-\frac{e^{\frac{\gamma}{2} t'}}{2}\big) \nonumber \\[1ex]
    & -i\int^{t}_{t_0}dt'e^{-i\omega_0 t'}e^{-\frac{\gamma}{2} t'}\Omega^*(t')\int^{t'}_{t_0}dt''e^{\gamma t''} \nonumber\\[1ex]
    &\left. \qquad \qquad \times \mathrm{Im}\left[\Omega (t'')\rho_{ge}(t'',t_0)\right]\right\} \,.\label{FormalCoh}
\end{align}
\end{subequations}
Finally, assuming the nucleus to be in its ground state initially, $\rho_{ee}(t_0)=0=\rho_{eg}(t_0)$, the self-consistent equations~(\ref{eq:SelfConsNoInit}) in the main text are obtained.

\section{Solution of the self-consistent equations\label{app:SelfDeriveSol}}

The self-consistent Eqs.~(\ref{eq:SelfConsNoInit}) can be expanded iteratively in orders of the (time-dependent) nucleus-field interaction $\Omega(t)$. In zeroth order, starting from an initial ground state, the solution is
\begin{subequations}
\begin{align}
    \rho_{ee}^{(0)}(t) &= 0\,, \\
    \rho_{ge}^{(0)}(t) &= 0\,.
\end{align}
\end{subequations}

In first order, inserting the zeroth-order solution into the right hand side of the self-consistent equations, one readily finds
\begin{subequations}
\label{eq:cons-single}
\begin{align}
     \rho_{ee}^{(1)}(t,t_0) &= 0\,, \\[2ex]
     \rho_{ge}^{(1)}(t,t_0) &= \int_{t_0}^{t}\: dt'\: f(t, t')\,, \\[2ex]
     f(t, t') &= -\frac{i}{2} e^{-\frac{\gamma}{2}(t-t')}e^{i\omega_0(t-t')}\: \Omega^*(t')\,. 
\end{align}
\end{subequations}
Iterating once more, we find for the second-order corrections
\begin{subequations}
\begin{align}
     \rho_{ge}^{(2)}(t,t_0) &= 0\,, \\[2ex]
     \rho_{ee}^{(2)}(t,t_0) &= 2\,\mathrm{Re}\left [ \int^{t}_{t_0}dt' \:f(t, t') \:  \int^{t'}_{t_0}dt'' \: f^*(t, t'') \right ] \nonumber \\[2ex]
     &= \left| \int^{t}_{t_0}dt' \: f(t, t')\right|^2 = \left |\rho_{ge}^{(1)}(t,t_0)\right|^2\,. \label{eq:equiv}
\end{align}
\end{subequations}
The first relation in Eq.~(\ref{eq:equiv}) follows from a simple re-arrangement of the different terms in the self-consistent Eq.~(\ref{eq:SelfPopNoInit}). In the second step, we have used a general relation for complex-valued functions. Note that $\rho_{ge}^{(2)}(t,t_0)$ in our notation is the second-order correction, and not the result up to second order.

\section{Solution of the $N$-body system\label{sec:ManyBodyProof}}

To derive the time-dependent coherences and populations of the $N$-body system in their respective leading order of the externally applied x-ray field, we rewrite the master equation Eq.~(\ref{eq:Master}) as
\begin{align}
\frac{d}{dt}\hat{\rho}^{\textrm{NB}} =& \frac{1}{i\hbar}\left[\hat{H}_0 +\hat{W}, \hat{\rho}^{\textrm{NB}}\right] \nonumber \\[1ex]
&- \left\{\hat{\Gamma}, \hat{\rho}^{\textrm{NB}}\right\} + \mathcal{L}'[\hat{\rho}^{\textrm{NB}}] \,, \label{master-rewritten}
\end{align}
where
\begin{subequations}
\begin{align} 
\hat{H}_0 &= \hbar \sum_{n=1}^N  \omega_0 \: \hat{\sigma}^+_n \hat{\sigma}^-_n -\hbar \sum_{n,n'=1}^N  J_{nn'} \: \hat{\sigma}^+_n \hat{\sigma}^-_{n'} \,,\\[2ex]
\hat{\Gamma} &=   \sum_{nn'}\left(\Gamma_{nn'}+ \delta_{nn'}\,\Gamma_{\mathrm{IC}}\right) \hat{\sigma}^+_{n}\hat{\sigma}^-_{n'}\,,\\[2ex]
\hat{W} &= - \frac{\hbar}{2} \sum_{n=1}^N \left[\Omega(\mathbf{r}_n,t) \,\hat{\sigma}^+_n +h.c.\right]\,, \\[2ex]
\mathcal{L}'[\hat{\rho}^{\textrm{NB}}] &= 2 \sum_{n ,n'=1}^N \left(\Gamma_{nn'}+ \delta_{nn'}\,\Gamma_{\mathrm{IC}}\right) \:  \hat{\sigma}^{-}_{n'}\hat{\rho}^{\textrm{NB}}\hat{\sigma}^{+}_{n}\,.
\end{align}
\end{subequations}
Note that we have written part of the Lindblad contribution as an anti-commutator $\left\{\cdot, \cdot \right\}$ with the density operator, for reasons which will become apparent later.

We aim at a perturbative expansion of the density operator in orders of $\hat W$ \cite{boyd2008nonlinear, Mukamel1995},
\begin{align} \label{eq:FormDensExp}
\hat{\rho}^{\textrm{NB}}(t) = \sum^{\infty}_{j=0} \hat{\rho}^{(j)}(t)
\end{align}
While treating $\hat{W}(t)$ perturbatively, the intrinsic nuclear and incoherent dynamics described by the master equation Eq.~(\ref{master-rewritten}) should be included nonperturbatively to all orders. For this, we employ an approach similar to the interaction picture used extensively in time-dependent perturbation theory in quantum mechanics and quantum field theory \cite{joachain1975quantum, sakurai1995modern,PeskinSchroeder1995} by defining the transformation operator
\begin{align}
\hat{T}_t &= \exp\left(-\hat{\Gamma}t+\frac{i}{\hbar}\hat{H}_{0}t\right )\,.
\end{align}
Note that $\hat{T}_t$ is not self-adjoint,
\begin{align}
    \hat{T}_t^\dagger &= \exp\left(-\hat{\Gamma}t-\frac{i}{\hbar}\hat{H}_{0}t\right )\,,\\[2ex]
    \hat{T}_t^{-1} &=  \hat{T}_{-t} = \exp\left(\hat{\Gamma}t-\frac{i}{\hbar}\hat{H}_{0}t\right )\,,    
\end{align}
because of the contribution of $\hat\Gamma$. For later reference, we note that $\hat{T}_t$ can be expressed as
\begin{subequations}
\label{eq:t-explicit}
\begin{align}
    \hat{T}_t &= \exp \left( \sum_{n,m} \: \notheta_{nm}\: \sigma_n^+ \sigma_m^- \: t\right) \,,\\[1ex]
    \notheta_{nm} &= -(\Gamma_{nm} + i\,J_{nm})  - (\Gamma_{\mathrm{IC}} - i \omega_{0})\delta_{nm}\,.
\end{align}
\end{subequations}
We note that the following derivation also applies to unequal two-level systems which differ in their respective single-particle decay rates $\Gamma_{nn}$ or transition frequencies $\omega_0$. However, since the desired equivalence between the coherently and incoherently scattered intensities in second order of the driving x-ray field  can only be established for identical two-level systems, we restrict the analysis to this case in the following. 

By rewriting the master equation in terms of 
\begin{align}
\hat{\rho}_I(t) &= \hat{T}_{-t}^{\dagger} \: \hat{\rho}^{\textrm{NB}}(t) \: \hat{T}_{-t} \label{eq:trafo} 
\end{align}
we obtain
\begin{align}
\dot{\hat{\rho}}_I(t) =& \frac{1}{i\hbar}\left[\hat{W}_I(t)\:\hat{\rho}_I-\hat{\rho}_I\:\hat{W}^{\dagger}_I(t)\right]\nonumber  \\[2ex]
&+ \hat{T}^{\dagger}_{-t} \: \mathcal{L}'[\hat{\rho}^{\textrm{NB}}] 
\: \hat{T}_{-t}, \label{eq:IntPicDynamics}
\end{align}
with the (non-Hermitian) transformed interaction part 
\begin{align}
\hat{W}_I(t) &= \hat{T}^{\dagger}_{-t}\hat{W}(t)\hat{T}^{\dagger}_t\,.
\end{align}

One can show, e.g., by explicit calculation, that the Lindblad contribution $\mathcal{L}'$ does not contribute in zeroth and first order in $\hat W$. In second order, it gives a contribution to the ground state part $|G\rangle\langle G|$ of the density operator, where $|G\rangle$ is the state with all 2-level systems in their ground state. This can be understood by noting that $\mathcal{L}'$ describes the feeding back of the spontaneous decay from the excited states into the respective ground states. The excited state population, however, only becomes non-zero in second order. Then, population decays into the ground state, but it can only be re-excited in higher-order of the expansion. As a result, we do not have to consider the $\mathcal{L}'$ contribution for our analysis, in which we are only interested in calculating the excited state populations and the coherences up to second order in $\hat W$. Hence, in the following, we neglect the $\mathcal{L}'$ contribution.

We continue by solving Eq.~(\ref{eq:IntPicDynamics}) by formal integration, followed by a transformation back to the original density operator,
\begin{align}
\hat{\rho}(t) &= \hat{T}^{\dagger}_{t-t_0}\, \hat\rho(t_0)\,  \hat{T}_{t-t_0} \nonumber \\[2ex]
&\quad -\frac{i}{\hbar} \int^{t}_{t_0}d\tau \:  \hat{T}^{\dagger}_{t-\tau}\left[\hat{W}(\tau),\hat{\rho}(\tau)\right]\hat{T}_{t-\tau} \,.
\end{align}

In lowest order in $\hat W$, we obtain
\begin{align}
\hat{\rho}^{(0)}(t) =&\, \hat{T}^{\dagger}_{t-t_0}\: \hat{\rho}^{(0)}(t_0)\: \hat{T}_{t-t_0} \,.
\end{align}
The first-order contribution becomes
\begin{align}
 \hat{\rho}^{(1)}(t) &=-\frac{i}{\hbar} \int^{t}_{t_0}d\tau \:  \hat{T}^{\dagger}_{t-\tau}\left[\hat{W}(\tau),\hat{\rho}^{(0)}(\tau)\right]\hat{T}_{t-\tau} \,.
\end{align}
Similarly, the second-order contribution is obtained by a further iteration,
\begin{align}
\hat{\rho}^{(2)}(t) =&\, - \frac{i}{\hbar}\int^{t}_{t_0}d\tau \: \hat{T}^{\dagger}_{t-\tau} \left[\hat{W}(\tau), \hat{\rho}^{(1)}(\tau)\right]\hat{T}_{t-\tau}  \,.
\end{align}
Using these expressions for the density operator, the relevant density matrix elements for each nuclear two-level system can be obtained as
\begin{subequations}
\begin{align}
\rho_{g_x,e_x}^{(1)}(t) &= \mathrm{Tr}\left[ \hat\sigma^+_x \, \hat{\rho}^{(1)}(t) \right]\,, \\[1ex]
\rho_{e_x,e_x}^{(2)}(t) &= \mathrm{Tr}\left[ \hat\sigma^+_x\,\hat\sigma_x^- \, \hat{\rho}^{(2)}(t) \right]\,.
\end{align}
\end{subequations}
One approach is to transfer to a basis in which the time evolution operator $\hat{T}_t$ becomes diagonal~\cite{lentrodt_ab_2020, Asenjo-Garcia2017}. Here, we instead proceed with a direct calculation in the single-particle basis.

We start with the coherence, assuming that all nuclei initially are in their ground states, $\hat\rho^{(0)}(t_0) = |G\rangle\langle G|$,
\begin{align}\label{eq-coh-1}
\rho_{g_x,e_x}^{(1)}(t) = &-\frac{i}{\hbar} \int^{t}_{t_0}d\tau \: \mathrm{Tr}\left[\hat \sigma^+_x \hat{T}^{\dagger}_{t-\tau}\left[\hat{W}(\tau),\hat{\rho}^{(0)}(\tau)\right]\hat{T}_{t-\tau} \right] \,.
\end{align}

The trace in the integral of Eq.~(\ref{eq-coh-1}) is evaluated to 
\begin{align}
\mathrm{Tr}&\left[\hat \sigma^+_x \hat{T}^{\dagger}_{t-\tau}\left[\hat{W}(\tau),\hat{\rho}^{(0)}(\tau)\right]\hat{T}_{t-\tau} \right] \nonumber \\[1ex]
&= \langle G \rvert \hat{T}_{t-\tau}\hat{\sigma}^+_x\hat{T}^{\dagger}_{t-\tau}\hat{W}(t)\lvert G\rangle -\langle G| \hat{W}(\tau)\hat{T}_{t-\tau}\hat\sigma^+_x \hat{T}^{\dagger}_{t-\tau}|G\rangle  \nonumber\\[1ex]
&= -\langle G| \hat{W}(\tau)\hat{T}_{t-\tau}\hat\sigma^+_x |G\rangle \label{eq:CohTrace}
\end{align}
where we have used the cyclic permutation property Tr($AB$) = Tr($BA$) in the first step, and 
\begin{align}
\hat{T}^{\dagger}_{t} |G\rangle = |G\rangle = \hat{T}_{t} |G\rangle \,,\\[1ex]
\langle G| \hat{T}_{t}  = \langle G| = \langle G| \hat{T}^\dagger_{t}\,,
\end{align}
as well as
\begin{align}
\langle G | \hat{\sigma}^+_x =0
\end{align}
in the second step. We further evaluate Eq.~(\ref{eq:CohTrace}) by inserting the explicit form of the interaction part $\hat{W}$ which yields
\begin{align}
 -\langle G| & \hat{W}(\tau)\hat{T}_{t-\tau}\hat\sigma^+_x |G\rangle \nonumber \\[1ex]
&= \frac{\hbar}{2} \sum_{n=1}^N \Omega^*(\mathbf{r}_n,\tau)\langle G| \hat{\sigma}^-_n\hat{T}_{t-\tau}\hat\sigma^+_x |G\rangle \nonumber \\
&= \frac{\hbar}{2} \sum_{n=1}^N \Omega^*(\mathbf{r}_n,\tau) \bigl[ \delta_{nx} + \notheta_{nx}(t-\tau)  \nonumber \\
& \qquad  + \sum_j \notheta_{nj}\notheta_{jx}(t-\tau)^2/2 +\dots\bigr]\nonumber \\
&= \frac{\hbar}{2} \sum_{n=1}^N \Omega^*(\mathbf{r}_n,\tau) \left[ e^{ \expmatrix(t-\tau)} \right]_{nx}\,. \label{eq:tmp1}
\end{align}
In the second step, we  perform a series expansion of the time evolution operator and calculate the expectation values. In the last step, we  re-sum the result. For this, we  introduce the matrix exponential of the coefficient matrix $\mathcal{K} = (\notheta_{nm})$ with entries $\notheta_{nm}$ defined in Eq.~(\ref{eq:t-explicit}), and $[A]_{xy}$ is the $x,y$-element of the matrix $A$. Using Eq.~(\ref{eq:tmp1}) in Eq.~(\ref{eq-coh-1}), we obtain
\begin{subequations}
\label{sol-coh}    
\begin{align}
\rho_{g_x,e_x}^{(1)}(t) &=   \sum_{n=1}^N \int^{t}_{t_0}d\tau \:  g_{nx}(\tau)\,,   \\[2ex]
g_{nx}(\tau) &=-\frac{i}{2} \Omega^*(\mathbf{r}_n,\tau) \left[ e^{\expmatrix(t-\tau)} \right]_{nx}\,. 
\end{align}
\end{subequations}

Analogously, the excited-state population evaluates to 
\begin{align}
\rho_{e_x,e_x}^{(2)}(t) &= \mathrm{Tr}\left[ \hat\sigma^+_x\,\hat\sigma_x^- \, \hat{\rho}^{(2)}(t) \right]\nonumber \\
&= \frac{2}{\hbar^2}\: \mathrm{Re}\left\{ \int_{t_0}^t d\tau_2 \int_{t_0}^{\tau_2}d\tau_1 \:\mathcal{M} \right\}\,, \label{pop2}
\end{align}
with the matrix element
\begin{align}
\mathcal{M} &= \langle G| \hat W(\tau_2) \hat T_{t-\tau_2} \hat\sigma_x^+ \hat\sigma_x^- \hat T^\dagger_{t-\tau_1} \hat W(\tau_1)|G\rangle  \nonumber \\[1ex]
&=  \frac{\hbar^2}{4}\sum_{nm} \Omega^*(\mathbf{r}_n,\tau_2) \Omega(\mathbf{r}_m,\tau_1)  \nonumber \\
&\qquad \times \langle G| \hat \sigma_n^- \hat T_{t-\tau_2} \hat\sigma_x^+ \hat\sigma_x^- \hat T^\dagger_{t-\tau_1} \hat\sigma_m^+|G\rangle \nonumber\\
&=  \frac{\hbar^2}{4}\sum_{nm} \Omega^*(\mathbf{r}_n,\tau_2) \Omega(\mathbf{r}_m,\tau_1)  \nonumber \\
&\qquad \times \sum_{\mathcal{N}} \langle G| \hat \sigma_n^- \hat T_{t-\tau_2} \hat\sigma_x^+  \,|\mathcal{N}\rangle \langle \mathcal{N}|\,\hat\sigma_x^- \hat T^\dagger_{t-\tau_1} \hat\sigma_m^+|G\rangle \nonumber\\
&=  \frac{\hbar^2}{4}\sum_{nm} \Omega^*(\mathbf{r}_n,\tau_2) \Omega(\mathbf{r}_m,\tau_1)  \nonumber \\
&\qquad \times \langle G| \hat \sigma_n^- \hat T_{t-\tau_2} \hat\sigma_x^+ |G\rangle \langle G| \hat \sigma_m^- \hat T_{t-\tau_1} \hat\sigma_x^+ |G\rangle^*\nonumber \\
&=  \frac{\hbar^2}{4}\sum_{nm} \Omega^*(\mathbf{r}_n,\tau_2) \Omega(\mathbf{r}_m,\tau_1)  \nonumber \\
&\qquad \times \left[ e^{\expmatrix(t-\tau_2)} \right]_{nx}  \left[ e^{\expmatrix(t-\tau_1)} \right]^*_{mx}\,. \label{eq-tmp-2}
\end{align}
In this derivation, the crucial step is the insertion of an identity operator $\hat{1}=\sum_{\mathcal{N}} |\mathcal{N} \rangle \langle \mathcal{N}|$ in the center of the matrix element where the $|\mathcal{N}\rangle$ form a  basis of the many-body Hilbert space. Of this sum, only the ground state projector $|G\rangle\langle G|$ contributes for the following reason: $\hat{\sigma}_m^+$ creates a single excitation in the state it is acting on, while the time evolution operator $\hat{T}^\dagger_{t-\tau_1}$ conserves the total number of excitations. Then, $\hat{\sigma}_x^-$ annihilates a single excitation such that only initial and final states with the same number of excitations can contribute to $\mathcal{M}$ which means that only the ground state projector remains in the center of the matrix element. 

By inserting Eq.~(\ref{eq-tmp-2}) into Eq.~(\ref{pop2}), we obtain
\begin{align}
\rho_{e_x,e_x}^{(2)}(t) &= \frac{1}{2}\: \mathrm{Re}\left\{ 
\int_{t_0}^t d\tau_2 \int_{t_0}^{\tau_2}d\tau_1 \sum_{nm} \Omega^*(\mathbf{r}_n,\tau_2) \Omega(\mathbf{r}_m,\tau_1)  \right.\nonumber \\
&\left. \qquad \times \left[ e^{ \expmatrix(t-\tau_2)} \right]_{nx}  \left[ e^{ \expmatrix(t-\tau_1)} \right]^*_{mx} \right\} \nonumber \\[2ex]
&= 2\: \mathrm{Re}\left\{ 
\int_{t_0}^t d\tau_2\: \sum_n \: g_{nx}(\tau_2)  \right. \nonumber \\
&\qquad \qquad \left. \times \int_{t_0}^{\tau_2}d\tau_1 \:\sum_m\: g^*_{mx}(\tau_1) \right\} \nonumber \\[2ex]
&= \left | \int_{t_0}^t d\tau_2\: \sum_n \: g_{nx}(\tau_2) \right|^2 \nonumber \\[2ex]
&=\left | \rho_{g_x, e_x}^{(1)}(t)\right|^2\,, \label{sol-pop}
\end{align}
where we have used Eqs.~(\ref{sol-coh}) and the general relation for complex-valued functions already employed in the derivation of Eq.~(\ref{eq:equiv}).

In summary, from Eqs.~(\ref{sol-coh}) and (\ref{sol-pop}) we thus obtain
\begin{subequations}
\label{sol-nbody}    
\begin{align}
\rho_{g_x,e_x}^{(1)}(t) &=   \sum_{n=1}^N \int^{t}_{t_0}d\tau \:  g_{nx}(\tau)\,,   \label{eq:nbody-coh}\\[2ex]
\rho_{e_x,e_x}^{(2)}(t) &= \left | \rho_{g_x, e_x}^{(1)}(t)\right|^2\,, \\[2ex]
g_{nx}(\tau) &=-\frac{i}{2} \Omega^*(\mathbf{r}_n,\tau) \left[ e^{ \expmatrix(t-\tau)} \right]_{nx}\,, \label{eq-g}
\end{align}
\end{subequations}
as the desired solution of the $N$-body dynamics.

As a consistency check, we can reduce the expression Eqs.~(\ref{eq:nbody-coh}) to the single-particle case. Then, the matrix exponential reduces to a scalar exponential with 

\begin{align}
\notheta_{xx} = -\Gamma_{xx} - \Gamma_{\mathrm{IC}} + i \omega_x= -\frac{\gamma}{2} + i \omega_0,
\end{align}
such that
\begin{align}
    \rho_{g_x,e_x}^{(1)}(t) &=  -\frac{i}{2}   \int^{t}_{t_0}d\tau \: \Omega^*(\tau) \: e^{ -\frac{\gamma}{2}(t-\tau)} \: e^{ i \omega_0(t-\tau)}\,,
\end{align}
which agrees with the single-particle result Eqs.~(\ref{eq:cons-single}).

\section{Nonresonant excitation of two-level systems beyond the low-excitation regime \label{sec:NonresAna}}

In Fig.~\ref{fig:FouLinComparison} we found that the \ffc spectra of the population exhibit diagonal structures with slope one, while the corresponding coherence squared may feature diagonals of slope one and two. In this Appendix, we explain this difference, based on the exact solution to a two-level system near-resonantly driven by a continuous light field with constant Rabi frequency $\Omega_0$~\cite{Scully1997}. 

The population and the coherence squared for this system can be written in terms of the generalized Rabi frequency $\Omega_{\Delta} = \sqrt{\Omega^2_0 +\Delta^2}$ as
\begin{align} 
\rho_{ee}(t) =& \frac{\Omega^2_0}{4\Omega^2_{\Delta}}\left(2-e^{i\Omega_{\Delta}t}-e^{-i\Omega_{\Delta}t}\right)\,, \nonumber\\[2ex]
\lvert\rho_{ge}(t) \rvert^2 =& \frac{\Omega^2_0}{16\Omega^4_{\Delta}}\left[6\Delta^2+2\Omega^2_{\Delta} - \Omega^2_0\left(e^{2i\Omega_{\Delta}t}+e^{-2i\Omega_{\Delta}t}\right) \right. \nonumber\\[2ex]
&\left.-4\Delta^2(e^{i\Omega_{\Delta}t}+e^{-i\Omega_{\Delta}t})\right]\,.
\end{align}
Here, we have rewritten the usual form of the expressions~\cite{Scully1997} in terms of exponential functions, since they directly reveal the different contributing frequency components. We find that the population features oscillations with the generalized Rabi frequency $\pm \Omega_{\Delta}$ only. In contrast, the absolute square of the coherence evolves with frequency $\pm \Omega_{\Delta}$ and $\pm 2\Omega_{\Delta}$. Note that the coherence $\rho_{ge}(t)$ itself does not comprise components oscillating at $\pm 2\Omega_{\Delta}$.  This suggests that the origin of the oscillation with double frequency in the coherence squared is a frequency-mixing between the negative and positive frequency component of the dipole oscillation.

In the limit $\Delta \gg \Omega_0$ of large detunings, $\Omega_{\Delta} \approx \Delta$, and thus oscillations with $\pm \Delta$ and $ \pm2\Delta$ appear that convert into the diagonal lines in the \ffc spectra in Fig.~\ref{fig:FouLinComparison} upon Fourier transformation along the time axis (cf.~\cite{PhysRevResearch.5.013071}). 

To demonstrate that the second pair of diagonal lines with slope two is a consequence of an excitation  beyond the \ler, we use the self-consistent Eqs.~(\ref{eq:SelfConsNoInit}) to derive the lowest and next-to-leading order results for the coherence squared. In order to focus on detuning-dependent effects, we again assume a pulse with constant envelope $\Omega_0$. The results read:
\begin{align}
\left(\lvert \rho_{ge}(t)\rvert^2\right)^{(0-2)} =& \frac{\Omega^2_0}{4\Delta^2}\left(2-e^{i\Delta t}-e^{-i\Delta t}\right)\label{eq:SecPopNoDecay}\,, \\[2ex]
\left(\lvert \rho_{ge}(t)\rvert^2\right)^{(4)} =&2\textrm{Re}\left[\rho^{*(1)}_{ge}(t) \rho^{(3)}_{ge}(t)\right] \\[1ex]
=& -\frac{7}{8}\frac{\Omega^4_0}{\Delta^4}+ \left(\frac{\Omega^4_0}{16\Delta^4}-i\frac{\Omega^4_0\,t}{8\Delta^3}\right)e^{i\Delta t} \nonumber \\[1ex]
&+\left(\frac{\Omega^4_0}{16\Delta^4}+i\frac{\Omega^4_0\,t}{8\Delta^3}\right)e^{-i\Delta t} \nonumber \\[1ex]
&-\frac{\Omega^4_0}{16\Delta^4}\left(e^{2i\Delta t} + e^{-2i\Delta t}\right)\,.
\end{align}
As expected from the general proof of the equivalence of population and coherence squared in second order of the x-ray-nucleus interaction, the lowest order contribution to the coherence squared features oscillations with $\pm \Delta$ only, like the population. In contrast,  already in fourth order the coherence squared oscillates with both frequencies $\pm \Delta$ and $\pm 2\Delta$. Therefore, we conclude that the diagonals with slope two only appear in the \ffc in case of excitation beyond the \ler. This suggests to use the appearance of the diagonal lines of slope two in the coherently scattered intensity as a feature to characterize dynamics beyond the \ler.


%

\end{document}